\documentclass[smallextended]{svjour3} 
\journalname{General Relativity and Gravitation}
\usepackage{graphicx}
\usepackage{amsmath,amssymb}
\usepackage{amsfonts}
\usepackage{time}
\usepackage{xspace} 
\usepackage[usenames]{color}
\usepackage{dcolumn}
\usepackage{bm}
\usepackage{mathrsfs}
\usepackage[colorlinks=true]{hyperref}
\usepackage[all]{hypcap} 
\usepackage[utf8]{inputenc}
\usepackage{multirow}
\usepackage{aas_macros} 
\usepackage{hyperref}
\usepackage{cite}
\usepackage{cancel}
\usepackage{authblk}
\usepackage{float}
\usepackage{subfigure}

\newcommand\be{\begin{equation}}
\newcommand\ba{\begin{eqnarray}}
\newcommand\ee{\end{equation}}
\newcommand\ea{\end{eqnarray}}

\begin{document}
\title{Love--$C$ relations for elastic hybrid stars}

\author{\textbf{Zoey Zhiyuan Dong}\textsuperscript{1,3} \and 
        \textbf{Joshua Cole Faggert}\textsuperscript{2,3} \and 
        \textbf{Shu Yan Lau}\textsuperscript{3} \and 
        \textbf{Kent Yagi}\textsuperscript{3}}
\authorrunning{Zoey Zhiyuan Dong \emph{et al}.}

\institute{
    \textsuperscript{1} College of Physical Science and Technology, Central China Normal University, Wuhan 430079, China \\
    \textsuperscript{2} School of Physics, Georgia Institute of Technology, Atlanta, Georgia 30332, USA \\
    \textsuperscript{3} Department of Physics, University of Virginia, Charlottesville, Virginia 22904, USA
}

\date{Received: date / Accepted: date}

\maketitle

\begin{abstract}

Neutron stars (NSs) provide a unique laboratory to study matter under extreme densities. Recent observations from gravitational and electromagnetic waves have enabled constraints on NS properties, such as tidal deformability (related to the tidal Love number) and stellar compactness. Although each of these two NS observables depends strongly on the stellar internal structure, the relation between them (called the Love--$C$ relation) is known to be equation-of-state insensitive. In this study, we investigate the effects of a possible crystalline phase in the core of hybrid stars (HSs) on the mass--radius and Love--$C$ relations, where HSs are a subclass of NS models with a quark matter core and a nuclear matter envelope with a sharp phase transition in between. We find that both the maximum mass and the corresponding radius increase as one increases the stiffness of the quark matter core controlled by the speed of sound, while the density discontinuity at the nuclear-quark matter transition effectively softens the equations of state. Deviations of the Love--$C$ relation for elastic HSs from that of fluid NSs become more pronounced with a larger shear modulus, lower transition pressure, and larger density gap and can be as large as 60\%.
These findings suggest a potential method for testing the existence of distinct phases within HSs, though deviations are not large enough to be detected with current measurements of the tidal deformability and compactness. 

\end{abstract}

\section{Introduction}

The state of cold matter at extremely high densities remains a major unresolved problem over the past decades. The lack of terrestrial experiments at relevant energy scales and the nonperturbative nature of the nuclear interactions form the main obstacles to obtaining a unified equation of state (EOS). Observations of neutron stars (NSs), therefore, provide an indirect but essential probe of the physics in this regime. 

Electromagnetic (EM) observations of pulsars allow one to probe the EOSs, focusing mainly on the mass, radius and spin properties. X-ray observations of millisecond pulsars provide independent constraints on the masses and radii through pulse profile modeling \cite{Bogdanov_2012, Ozel_2016, Miller_2019,Riley:2019yda, Miller_2021,Riley:2021pdl}. Radio pulsar timing also constrains the EOS through the mass measurements (e.g., \cite{Cromartie_2020}). 

Gravitational wave (GW) observations of binary NS coalescences can constrain EOSs from the tidal deformability measurement. 
This was demonstrated for the first binary NS merger event, GW170817 \cite{Abbott_2017a, LIGOScientific:2018hze, Annala_2018}. The EM counterparts, including the short gamma-ray burst event, GRB 170817A, and the astronomical transient, AT2017gfo, also opened up multimessenger astronomical analysis of the same source \cite{Abbott_2017b, Radice_2018, Dietrich_2020}.  With the addition of the next-generation GW detectors in the coming decades, the detection horizon of binary NS mergers extends to the redshift of $z\sim3$, and more than $10^4$ events are expected to be found per year with moderate to high signal-to-noise ratios \cite{Maggiore_2020, Luck_2021}. The finite-size effects of NSs can be probed not only from the late inspiral phase but also from the postmerger signal. This offers further insights into the internal structure of NSs through, e.g., the stellar oscillation modes \cite{Williams_2022}, and hence leading to stronger constraints on the EOS within the pressure-density plane \cite{Breschi_2022, Finstad_2022}.

The EOS of the core of an NS is the most uncertain region due to the various possible phases of matter emerging, like pions, hyperons, and deconfined quarks \cite{Shapiro_1983, Haensel_2007}. In particular, deconfinement is a consequence of the asymptotic freedom in quantum chromodynamics (QCD). An astrophysical compact object with these deconfined quarks inside its core is called a hybrid star (HS), featuring a transition region between the quark matter (QM) core and nuclear matter (NM) envelope.

How the transition between NM and QM occurs is still uncertain.
The transition can have a density discontinuity across the interface between the two phases, known as the sharp transition scenario, following the Maxwell construction. In contrast, Glendenning \cite{Glendenning_1992} proposed a type of soft transition through the Gibbs construction, leaving a mixed phase between the NM and QM phases, which smoothens the density profile of an HS. Both scenarios correspond to a self-consistent treatment of a first-order phase transition, imposing different equilibrium conditions that depend on the surface tension between the two phases (e.g., \cite{Alford_2001}). Recent work has also considered smooth crossover transitions \cite{Masuda_2013, Baym_2018, Han_2019} constructed from an interpolation between the two phases without assuming the equilibrium conditions, which also imply a mixed phase between the NM and QM phases. While the first-order phase transitions (either soft or sharp) soften the EOS in general, the crossover transition causes stiffening if the QM part is stiff enough \cite{Masuda_2013}. Although the true nature of the transition is poorly known, we focus on the sharp phase transition scenario in the following, where the density discontinuity is expected to impact the HS properties.

Various scenarios have been proposed on HSs with solid QM cores.
One possibility is the crystalline color superconducting (CCS) phase \cite{Casalbuoni_2002, Mannarelli_2006, Rajagopal_2006, Mannarelli_2007}, in which Cooper pairs of color charges attain non-zero momenta and form a condensate with broken translational and rotational symmetries, i.e., a crystal. In condensed matter conventions, this is known as a ``LOFF" state \cite{Larkin_1964, Fulde_1964}. Another possibility has been discussed in \cite{Xu_2003}, where the QM solidifies by forming ``quark clusters" \cite{Michel_1988}. In this paper, we define HSs with a solid core and a fluid envelope as \emph{elastic} HSs, while those with a fluid core are termed \emph{fluid} HSs.

The presence of elasticity is known to have an impact on observables, such as tidal deformability. Lau \emph{et al}. \cite{Lau_2017} showed that a quark star composed entirely of solid matter could have tidal deformability 60\% lower than a perfect fluid quark star (see also \cite{Lin_2007, Lau_2019} for related works). Pereira \emph{et al}. \cite{Pereira_2020} computed the tidal deformability of HS models with a solid layer, either from the NM crust or the mixed phase of the NM-QM phase transition. They demonstrated that the tidal deformability can change by $\sim 5\%$ from the NS case if the thickness of the solid layer is more than half of the stellar radius, and this amount of change may be detectable with future GW observations. Previous work has also studied the effect of elastic HSs on asteroseismology \cite{Lin_2013, Mannarelli_2014, Mannarelli_2015}.

The Love--$C$ relation, relating the tidal deformability (related to the tidal Love number) and the compactness ($C$) in NSs and HSs, are known to be EOS-insensitive~\cite{Maselli_2013, Yagi_2017, Jiang:2020uvb}. Many other universal relations are known to exist, but the Love--$C$ relation is particularly interesting from observational viewpoint as the tidal deformability has been measured through GW observations with LIGO/Virgo (GW170817 \cite{Abbott_2018}) while the compactness has been measured through X-ray observations with NICER and XMM-Newton (PSR J0030+0451 \cite{Miller_2019} and PSR J0740+6620 \cite{Miller_2021}).
 In \cite{Yagi_2013, Yagi_2017, Carson_2019}, fitting formulas are provided for NSs and HSs with sharp phase transitions, where the fractional deviations of the various EOSs considered are all within 10\%. Such approximate universal relations are particularly useful in inferring the compact star properties through NS observations~\cite{Yagi_2013, Yagi2013_b, Yagi_2017, Abbott_2018}. For instance, it provides a way to estimate the radius from a simultaneous measurement of the mass and tidal deformability, as in \cite{Abbott_2018}. Alternatively, simultaneous measurements of both the tidal deformability and the compactness allow us to test the nature of astrophysical compact stars. 

\begin{figure}[H]
\includegraphics[width=6.cm]{ 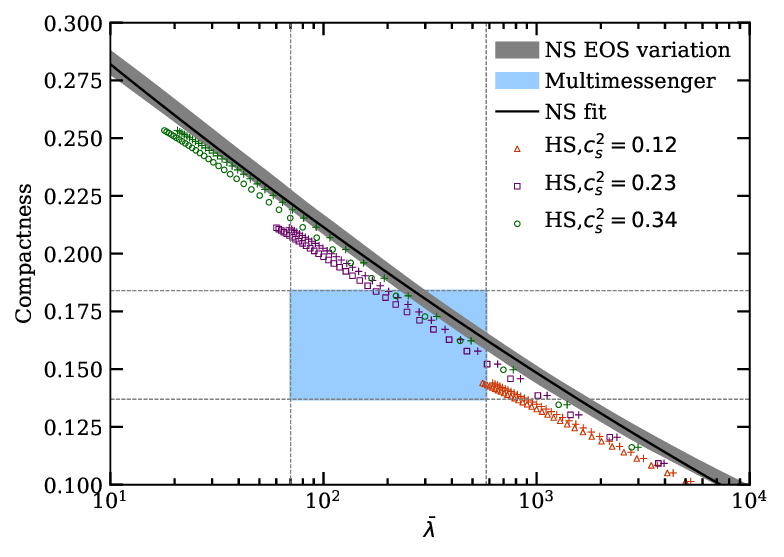}
\includegraphics[width=6.cm]{ 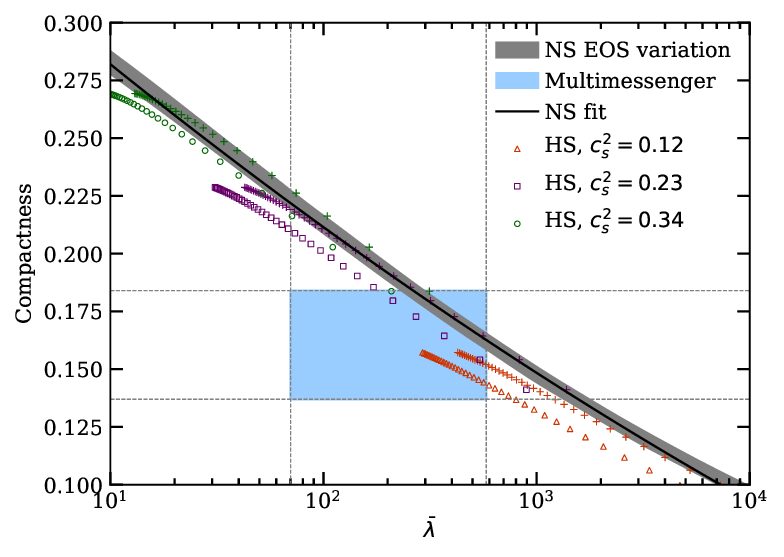}
\caption{\label{fig:LoveC_vary_cs2} 
Love--$C$ relations for HSs with the varied speed of sound $c_s$. We fix the transition pressure as $P_t = 2\times10^{34}~\text{dyn cm}^{-2}$ (left) and $P_t = 2\times10^{33}~\text{dyn cm}^{-2}$ (right) while we set the energy density discontinuity as $\Delta \rho = 1\times10^{14}~\text{g cm}^{-3}$. NM is assumed to follow the APR EOS. We present the results for both HSs with solid (open symbols) and fluid (crosses) QM cores.  We also present the fit (Eq.~\eqref{eq:love_c_fit}) for the Love--$C$ relation for fluid NSs in \cite{Yagi_2017} (solid line). The EOS variation for this relation is taken from \cite{Postnikov_2001} (grey-shaded region). The measurement uncertainties on the stellar compactness from NICER \cite{Silva_2021} and tidal deformability from LIGO/Virgo \cite{Abbott_2017a} for a star with 1.4$M_\odot$ are also overlaid (blue shaded region). }
\end{figure}

In this paper, we study elasticity within HSs on the Love--$C$ relation. Given the uncertainty in the NM-QM transition, an HS could contain a thick, solid QM core that can induce a significant difference in the Love--$C$ relation from the NS case. We start by constructing static spherically symmetric background solutions of HSs using a parametrized EOS model for the QM and realistic EOS tables for NM, assuming the background is unstrained. We then statically perturb the star, including the elasticity effect in the QM phase, to obtain the tidal deformability. Our result demonstrates that the elasticity causes substantial deviations in the Love--$C$ relation from the NS relation. 
However, the deviations are unfortunately not large enough to be distinguishable by the current measurement uncertainties of the tidal deformability and compactness from LIGO/Virgo and NICER/XMM-Newton. Our main findings are summarized in  Fig.~\ref{fig:LoveC_vary_cs2}.

The rest of this paper is organized as follows: In Sec.~\ref{sec:background}, we describe the construction of the equilibrium HS models. Then, we briefly describe the formalism to compute the tidal deformability in Sec.~\ref{sec:tidal_deformability}. We present our numerical results in Sec.~\ref{sec:results}. Finally, we provide a summary in Sec.~\ref{sec:conclusion}. Unless otherwise specified, we use the geometrized unit system with $G = c =1$.

\section{Equilibrium background}
\label{sec:background}

To calculate the structure and the relativistic tidal deformability of elastic HSs, we start by solving the Tolman-Oppenheimer-Volkoff (TOV) equations for the static spherically symmetric background \cite{PhysRev.55.374, PhysRev.55.364}. We assume the background configuration is unstrained so that the effect of elasticity does not affect the background solution. The method is standard, and we provide a brief review below. 

\subsection{TOV equations}

We begin by providing the line element for a spherically symmetric spacetime in the standard Schwarzschild coordinates:
\begin{align}
    ds^2 = -e^{\nu(r)} dt^2 + e^{u(r)} dr^2 + r^2 d\theta^2 + r^2 \sin^2\theta d\phi^2.
\end{align}
One can derive the TOV equations by plugging in the above metric ansatz to the Einstein equations: 
\begin{align}
    \nu^\prime &= 2\frac{m+4\pi r^3 p}{r^2} e^u, \label{eq_TOV1}\\
    p^\prime &= -(\rho+p)\frac{\nu^\prime}{2},\label{eq_TOV2}\\
    u^\prime &= 2\frac{-m+4\pi r^3 \rho}{r^2} e^u, \label{eq_TOV3}\\
    e^u &= \left(1-\frac{2m}{r}\right)^{-1},\label{eq_TOV4}
\end{align}
where $p$ is the pressure, $\rho$ is the energy density, $m$ is the gravitational mass contained within a sphere of radius $r$, and the prime symbol denotes the derivative with respect to $r$. The quantities $\nu$, $u$, $m$, $p$, $\rho$ are all functions of $r$.
The pressure and the energy density are related through the EOS, which gives $\rho = \rho(p)$ in the static equilibrium background (see Sec.~\ref{ssec:EOS} for more details). 

We can numerically solve the above TOV equations as follows. By choosing a central value of the pressure $p_c$ or energy density $\rho_c$, we integrate the equations at some small $r$ with the following boundary condition:
\begin{equation}
    p = p_c + \mathcal{O}(r^2), \quad \nu = \nu_0 r^2 + \mathcal{O}(r^4), \quad m = \frac{4\pi}{3} \rho_c r^3 + \mathcal{O}(r^5)\,.
\end{equation}
Here, $\nu_0$ is initially an arbitrary constant that will be fixed with another boundary condition
\begin{equation}
    \nu(R) = \ln\left(1-\frac{2M}{R} \right).
\end{equation}
Here, $R$ is the stellar radius determined by the condition $p(R) = 10^{-12}p_c$ while $M = m(R)$ is the stellar gravitational mass.
This allows us to determine the static profile $p(r)$, $m(r)$, $\nu(r)$ for each EOS with a chosen central pressure or density. In particular, we obtain the compactness of a star, defined by 
\begin{align}
    C = \frac{M}{R}.
\end{align}

\subsection{Equation of state}
\label{ssec:EOS}

To obtain the solution of the Einstein equations, we need the EOS that relates the thermodynamic quantities. 
For cold NSs at equilibrium, the EOS is given as a relation between $\rho$ and $p$. For the NM component, we employ the tabulated EOS APR \cite{Akmal_1998}. In Appendix~\ref{NL3}, we also use a tabulated EOS NL3 \cite{NL3_EOS, NL3_EOS_b} and MPa \cite{PhysRevC.90.045805} for reference. For the QM region, we employ a parametrized model known as the constant speed of sound (CSS) model \cite{Alford_2013}. This model is inspired by the results from perturbative QCD in the asymptotic regime, where the speed of sound of QM is almost independent of the density. 
The resulting EOS is given by \cite{Alford_2013}
\begin{align}
\rho(p) = 
\begin{cases}
       \frac{p - P_t}{c_s^2} + \rho_\text{NM}(P_t) + \Delta\rho,  &\text{if } p > P_t\\
       \rho_\text{NM}(p),   &\text{if } p \leq P_t,
\end{cases}
\end{align}
where $\rho_\text{NM}(p)$ is the NM EOS obtained by a log-linear interpolation from the corresponding EOS table, $c_s$ is the speed of sound that is assumed to be constant, $P_t$ is the transition pressure, and $\Delta \rho$ is the energy density gap between NM and QM. We treat $c_s^2$, $P_t$, and $\Delta \rho$ as the model parameters, with $c_s^2 \leq 1$, and all these parameters are arbitrary (see \cite{Alford_2013} for the effect of varying $c_s$ and $\Delta \rho$ on the HS mass-radius relations).
The QM and NM EOSs are joined by assuming a continuous pressure to obtain a sharp phase transition that contains a density gap. 

The theoretical bounds on the CSS parameters deserve some clarifications. The CSS model can be considered as the leading order expansion of the QM EOS in the high-density limit. A similar and well-known EOS is the MIT Bag model \cite{Chodos_1974}, which describes massless, non-interacting quarks and is related to the CSS parametrization with $c_s^2 = 1/3$. This model contains a phenomenological parameter, the `bag constant', to account for the nonperturbative effects that give rise to confinement. This effectively describes the vacuum pressure exerted on the QCD vacuum. The range of this parameter is highly uncertain. If we assume the strange matter hypothesis to hold, i.e., a matter formed from up, down, and strange quarks is absolutely stable, the value of the bag constant then lies between 58.9 to 91.8 $\text{MeV fm}^{-3} $\cite{Bodmer_1971, Witten_1984, Haensel_2007}. 
In the CSS model, the effect of this vacuum pressure is encapsulated in $\Delta \rho$, and one can restrict the CSS parameters through this theoretical bound. 
However, the mapping between $\Delta \rho$ and the MIT bag model depends on additional parameters related to the quark interactions other than just the bag constant (see, e.g., the quartic term in \cite{Alford_2005}). In addition, variants of the bag model, like the quadratic terms introduced in \cite{Alford_2005} or density-dependent bag constants \cite{Jin_1997, Burgio_2002}, further widen the range of the theoretical constraints on $c_s^2$ and $\Delta \rho$.
Therefore, We allow the CSS parameters to vary in a wide range of values.

\begin{figure}[H]\centering
\includegraphics[width=6.cm]{ 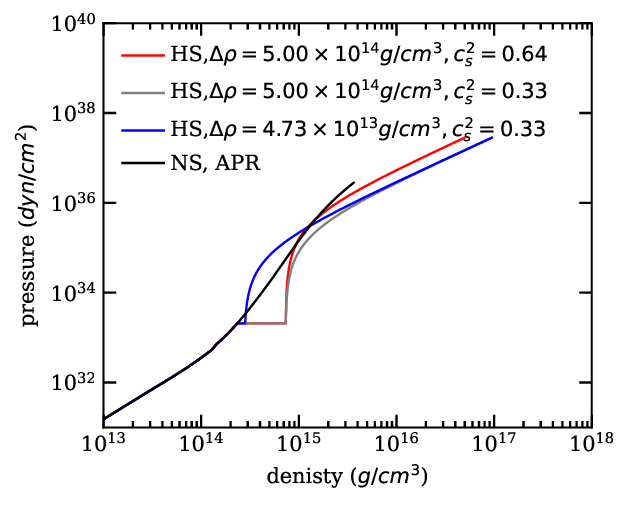}
\caption{\label{fig:EOS example} The EOSs for NSs and HSs. The pressure-density relation is shown for the NS APR EOS (black) and for HSs (colored) with different values of the speed of sound ($c_s^2$) and the energy density gap ($\Delta \rho$). 
The transition pressure is fixed at $2\times10^{33}~\text{dyn cm}^{-2}$.}
\end{figure}

Figure~\ref{fig:EOS example} illustrates several examples of EOSs for different values of the speed of sound ($c_s^2$) and the density gap ($\Delta \rho$) at a transition pressure $P_t=2 \times 10^{33} ~\text{dyn cm}^{-2}$. The APR EOS, shown in a black solid line, represents the baseline NS model, while the HS models, depicted in red, grey, and blue, highlight the effects of varying the speed of sound and density gap on the pressure-density relation. We will discuss the implications of these variations on the structure of HSs in detail in Sec.~\ref{ssec:MR_constrants}.

\subsection{Shear modulus model}

We next describe the EOS of the shear modulus $\mu$ for the QM phase used in this paper. This quantity measures the material’s ability to resist shearing, affecting how NSs deform in response to tidal forces. Its relation with the stress and strain variables is described in Sec.~\ref{sec:tidal_deformability}.
Here, we adopt the following parametrized model 
\begin{align}
    \mu = \mu_\mathrm{ref} \left(\frac{\rho}{\rho_\text{ref}}\right)^{N}, \label{eq:shear_modulus}
\end{align}
where $\mu_\mathrm{ref}$ is the shear modulus at a reference energy density that we set as  $\rho_\text{ref} = 2\times10^{15}~\text{g cm}^{-3}$ while the index $N$ characterizes the energy-density dependence of the shear modulus.

The above power-law relation is inspired by the shear modulus of the CCS phase derived in \cite{Mannarelli_2007}, given by
\begin{align}
    \mu_\text{CCS} = 2.47~\text{MeV fm}^{-3} \left(\frac{\Delta}{10 ~\text{MeV}}\right)^2\left(\frac{\mu_q}{400 ~\text{MeV}}\right)^2,
\end{align}
where $\mu_\text{q}$ is the quark chemical potential, and $\Delta$ is the gap parameter of the CCS phase, which is estimated to range from 5 MeV to 25~$\text{MeV}$. In the high-density limit, $\mu_\text{CCS} \propto \mu_q^2$, and $\rho \propto \mu_q^4$ as for the ultrarelativistic free Fermi gas. Hence, we have $ \mu_\text{CCS} \propto \sqrt{\rho}$, which corresponds to $N = 0.5$ in Eq.~\eqref{eq:shear_modulus}.

\section{Tidal deformability} \label{sec:tidal_deformability}

We now describe how to compute the tidal deformability of HSs.
A tidally deformed star can be constructed by solving the perturbed Einstein equations, which relate the perturbed metric to the perturbed stress-energy tensor. The even-parity time-independent metric perturbation in the Regge-Wheeler gauge \cite{Regge_1957} is written as
\begin{align}
    \delta g_{ab} dx^a dx^b = \sum_{l,m} \Big[&-e^\nu H_0(r) dt^2 + e^{u} H_2(r) dr^2  \nonumber\\
    &+  r^2 K(r) \left(d\theta^2 + \sin^2 \theta d\phi^2\right)\Big] Y_{lm}(\theta,\phi),
\end{align}
where $Y_{lm}$ is the spherical harmonics. Hereafter, we consider modes of perturbations with specific $l$ and $m$, particularly on the $l=2$ perturbations for the tidal problem.

On the other hand, the perturbed stress-energy tensor contains the contribution from elasticity:
\begin{align}
    \delta T_ {a b} = \delta T^\text{bulk}_{a b} + \delta T^\text{shear}_{a b},
\end{align}
where
\begin{align}
    \delta T^\text{bulk}_{a b} &= \delta (\rho u_a u_b + p h_{ab}), \\
    \delta T^\text{shear}_{a b} &= -2\mu \delta \Sigma_{a b}. \label{eq:Hookes}
\end{align}
Here, $\mu$ is the shear modulus, $u^a$ is the four-velocity of the bulk, $h_{ab} = g_{ab} + u_a u_b$ is the projection tensor perpendicular to $u^a$, $\delta X$ denotes an Eulerian perturbation of the physical quantity $X$. $\delta \Sigma_{a b}$ is the Eulerian perturbation of the constant-volume strain tensor (shear tensor), which is the same as its Lagrangian perturbation $\Delta \Sigma_{a b}$ under the assumption that the background is unstrained and is given by \cite{Carter_1972, Penner_2011, Pereira_2020, Gittins_2020, Andersson_2021}
\begin{align}
    \delta \Sigma_{a b} =& \Delta\Sigma_{a b}\nonumber\\
    =& \frac{1}{2}\left(h^c_a h^d_b - \frac{1}{3}h_{ab}h^{cd}\right)\Delta g_{cd}.
\label{eq:strain_tensor}
\end{align}
Here the Lagrangian perturbation of the metric $g_{ab}$ is given by \cite{Friedman_1975}
\begin{align}
    \Delta g_{ab} = \delta g_{ab} + \nabla_{a} \xi_{b}+\nabla_{b} \xi_{a}.
\end{align}
The symbol $\nabla_{a}$ is the covariant derivative, while the (even-parity) Lagrangian displacement vector components are 
\begin{align}
    \xi^r =& \sum_{l,m} \frac{W(r)}{r} Y_{lm}(\theta,\phi),\\
    \xi_A =& \sum_{l,m} V(r) \partial_A Y_{lm}(\theta,\phi),
\end{align}
where the index $A$ runs between $\theta$ and $\phi$. The perturbations $\delta\rho$ and $\delta p$ are also expanded in terms of spherical harmonics. 
For static perturbations, 
\begin{align}
    \delta \rho = \frac{d\rho}{dp} \delta p = \frac{\delta p}{c_s^2}. \label{eq:cs}
\end{align}
Note that $\delta u_a = \delta (g_{ab} u^b) = \delta g_{ab} u^b + g_{ab} \delta u^b$.

The above equations allow us to construct a set of six coupled ordinary differential equations (ODEs) that govern the perturbed solid core. The perturbations in the fluid envelope can be obtained by setting $\mu = 0$. We follow the formalism in \cite{Lau_2019} for the ODEs in the solid core (which is consistent with the one in \cite{Pereira_2020, Gittins_2020} but written in terms of different variables) and the formalism in \cite{Hinderer_2008, Hinderer_2009} for the fluid part. 
We follow the formalism and method in \cite{Lau_2019} for solving the ODEs. See Appendices~\ref{sec_formalism} and \ref{Computation method} for more details, including the procedure for deriving the tidal deformability of elastic HSs and the computation method.

Solving the boundary value problem allows us to determine the tidal deformability, $\lambda$, defined as 
\begin{align}
    Q_{ij} = -\lambda \mathcal{E}_{ij},
\end{align}
where the quadrupole moment $Q_{ij}$ and the external tidal field $\mathcal{E}_{ij}$ are defined from the asymptotic behavior of the $(t,t)$ component of the metric \cite{Thorne_1998}:
\begin{align}
    g_{tt} = -1+\frac{2 M}{r} &+ 3 \frac{Q_{ij}}{r^3} \left(\frac{x^ix^j}{r^2} - \frac{1}{3}\delta^{ij}\right) \nonumber \\
    &+ \mathcal{O}\left(\frac{R^4}{r^4}\right) - \mathcal{E}_{ij}x^ix^j + \mathcal{O}\left(\frac{r^3}{L^3}\right),
\end{align}
where $L \gg r \gg R$, with $L$ being the length scale
associated with the radius of curvature from the external gravitational source, $x^i$ is the position vector in Cartesian coordinates, and $\delta^{ij}$ is the Kronecker delta symbol.
The tidal deformability quantifies how much a celestial body deforms in response to the external tidal field.
In the following, we also introduce a dimensionless tidal deformability, $\bar\lambda = \lambda/M^5$ \cite{Yagi_2013}, when studying universal relations.

Before we discuss the results, here is a caveat about the QM-NM interface under perturbation. In our formalism, we have employed the ``slow conversion condition" between the solid core and fluid envelope, which assumes the two phases at the two sides of the interface stay intact after perturbations \cite{Karlovini_2004, Pereira_2018, Pereira_2020}. Since the tidal perturbations occur in finite timescales set by the process (e.g., the inspiral timescale), the boundary condition at the interface depends on the phase conversion rate against the perturbation timescale. 
The slow conversion condition assumes that the conversion is much slower than the perturbation process.
In another limit where the phase conversion between the two phases is much faster than the perturbation timescale, a different boundary condition should be used. 
The actual boundary condition may lie between these two limits and depend on the QM-NM conversion rate.
We discuss it briefly in Appendix~\ref{app:rapid_bc} and leave further investigation as future work.

\section{Numerical results}\label{sec:results}

In this section, we present the $M$--$R$ relations and Love--$C$ relations for elastic HSs by varying the EOS parameters, which include $c_s$, $\Delta \rho$, and $N$. 

\subsection{\label{ssec:MR_constrants}Mass-radius relations}

First, we consider the existing observational constraints on the mass and radius of the equilibrium background of the elastic HS models where the NM part employs APR (see Appendix \ref{NL3} for NL3 and MPa). We numerically construct the models as described in Sec.~\ref{sec:background} to obtain the $M$--$R$ relations for a set of EOS parameters.

\subsubsection{Dependence on the stiffness of quark matter}

Let us first focus on varying $c_s$. The speed of sound quantifies the stiffness of the QM core, which plays an essential role in determining the mass and size of the HS. However, the speed of sound of QM is highly uncertain within the high-density and nonperturbative region. Perturbative QCD predicts that $c_s^2$ approaches the free relativistic gas limit of $1/3$ from below, known as the conformal limit \cite{Bedaque_2015}. Despite this, there is no guarantee that the NS or HS interior cannot exceed this limit \cite{Fujimoto_2022}. Therefore, we consider QM within the constant sound speed model with $0 < c_s^2 \le 1$, with the upper bound set by the causality limit, while observations of high-mass NSs restrict $c_s^2$ from below. 

\begin{figure}[H]
\includegraphics[width=6.cm]{ 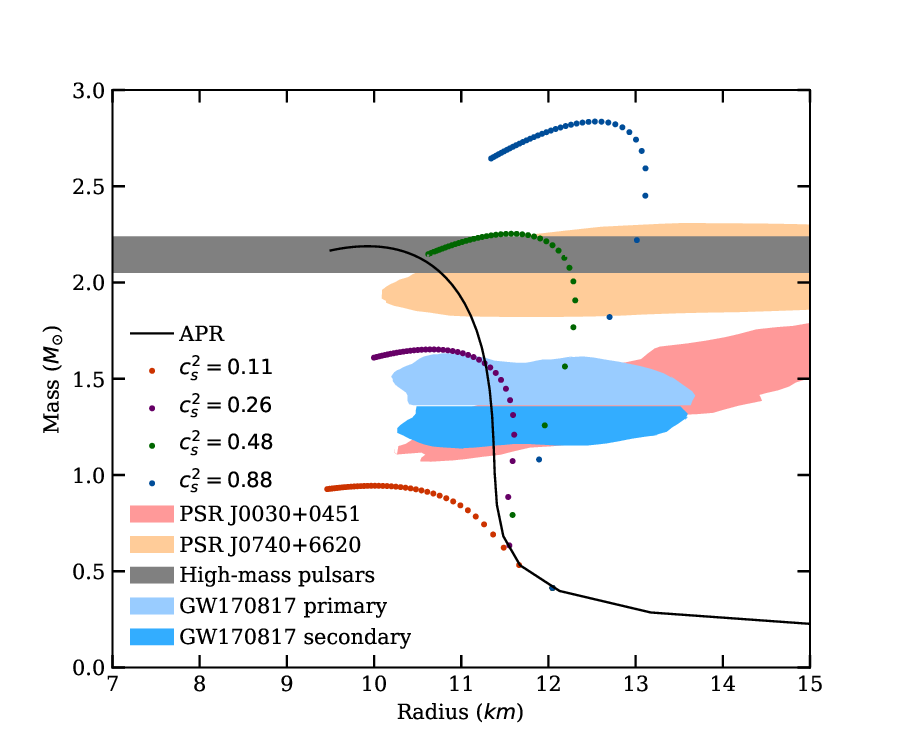}
\includegraphics[width=6.cm]{ 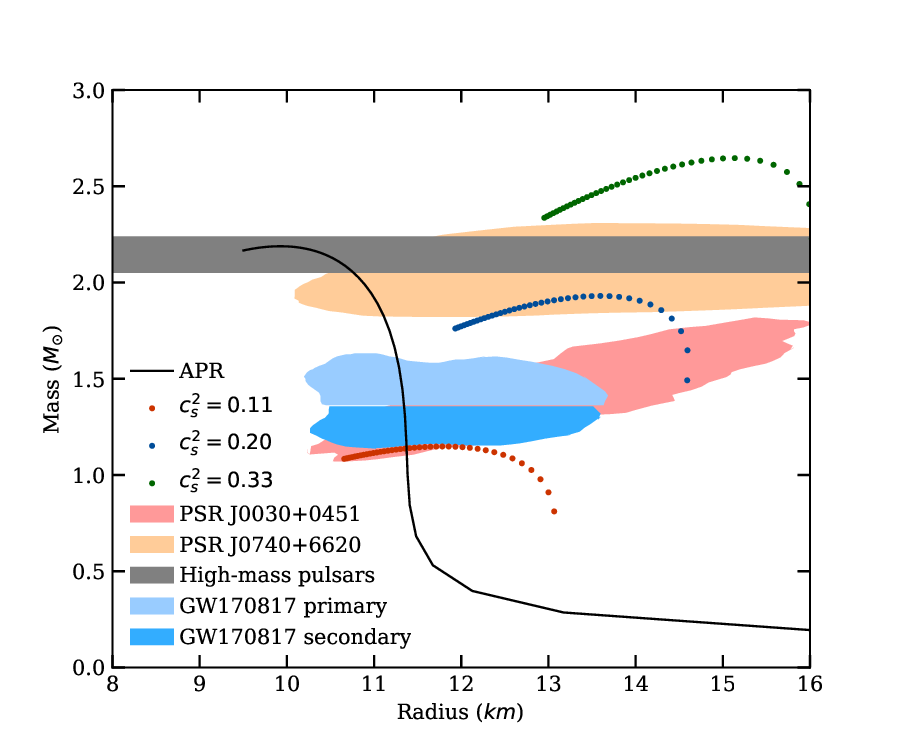}
\caption{\label{fig:MR_vary_cs2} $M$--$R$ relations for the HS models constructed with the EOS described in Sec.~\ref{ssec:EOS}. We vary $c_s^2$ while fixing $P_t = 2\times10^{34}~\text{dyn cm}^{-2}$ (left) and $P_t = 2\times10^{33}~\text{dyn cm}^{-2}$ (right). We choose $\Delta \rho = 5\times10^{12}~\text{g cm}^{-3}$ while using APR as the NM EOS. 
We also present the relation for NSs with APR as the NM EOS. Various observation bounds are shown with shaded regions, including the mass measurement of J0740+6620 from pulsar timing \cite{Cromartie_2020} (68.3\% confidence interval, grey horizontal band), and the mass and radius measurement of GW170817 \cite{Abbott_2018} (90\% confidence interval, orange and blue regions separated by a horizontal blue line), J0030+0451 with NICER (95\% confidence interval, maroon region) \cite{Miller_2019}, and J0740+6620 with NICER and XMM-Newton (95\% confidence interval, indigo region) \cite{Miller_2021}.
}
\end{figure}

In Fig.~\ref{fig:MR_vary_cs2}, we show the $M$--$R$ relations for two sets of EOSs with $\Delta \rho = 5\times10^{12}~\text{g cm}^{-3}$ and $P_t$ at $2\times10^{34}~\text{dyn cm}^{-2}$ (left panel) and $2\times10^{33}~\text{dyn cm}^{-2}$ (right panel) respectively. These values of $\Delta \rho$ and $P_t$ are just reference values and are chosen somewhat arbitrarily (See Sec.~\ref{ssec:EOS} for detailed explanation). We choose APR for the NM EOS and vary the stiffness of the QM phase by increasing $c_s^2$ from $0.11$ to $0.88$. Note that since we assume that the core at equilibrium is unstrained, the mass-radius relations are not affected by the presence of a solid core. The numerical results demonstrate that HS EOSs support a larger maximum mass and radius as we increase the stiffness of the core. This is evident from Fig.~\ref{fig:EOS example}, where the value of $c_s^2$ corresponds to the slope of the EOS function in the core region on a linear scale. With a higher value of $c_s^2$, as represented by the red line, the pressure corresponding to a given density is higher. The effect is larger for the lower $P_t$ case due to the larger size of the QM core.

In the same figure, we also overlay the observational constraints on the $M$--$R$ relations. 
These include the mass measurement of PSR J0740+6620 \cite{Cromartie_2020}, as well as inferred mass and radius bounds from the binary NS merger event GW170817 \cite{Abbott_2018}, PSR J0030+0451 \cite{Miller_2019} with NICER, and PSR J0740+6620 \cite{Miller_2021} with NICER and XMM-Newton. These constraints, in particular PSR J0740+6620, rule out part of the soft HS EOSs with a smaller $c_s$ for higher $P_t$. For lower $P_t$, the HS EOSs are less likely to satisfy the observation bounds, either the maximum mass being too low or the radius being too large. These results suggest that for a relatively small $\Delta \rho$, lower $P_t$ is not favoured, and $c_s$ has to be large for higher $P_t$ due to the $M$--$R$ constraints.

\subsubsection{Dependence on the energy density discontinuity}

\begin{figure}[H]
\includegraphics[width=6.cm]{ 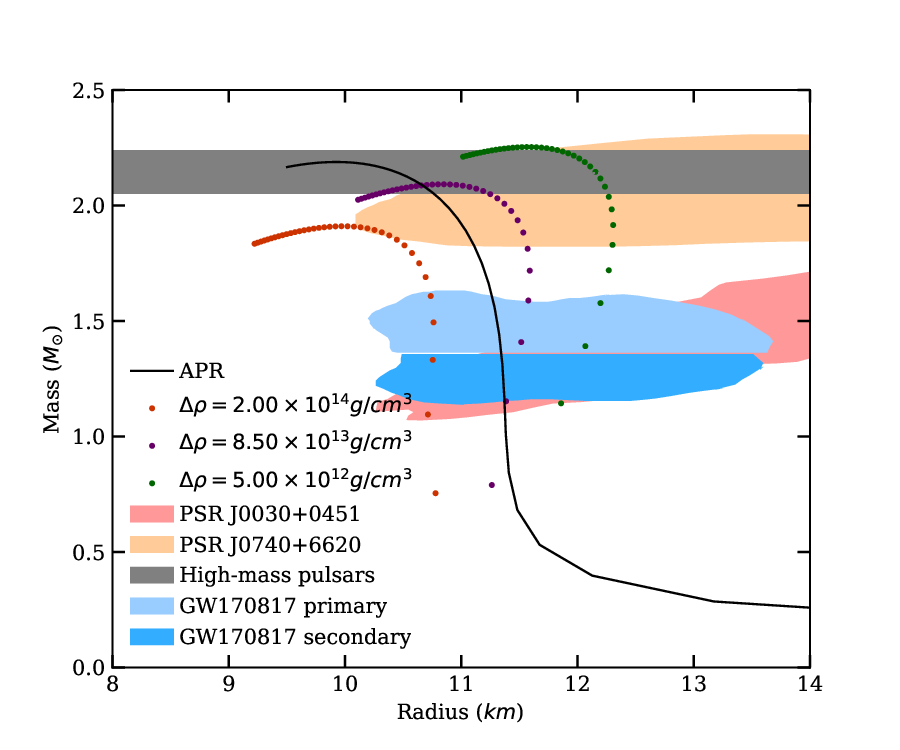}
\includegraphics[width=6.cm]{ 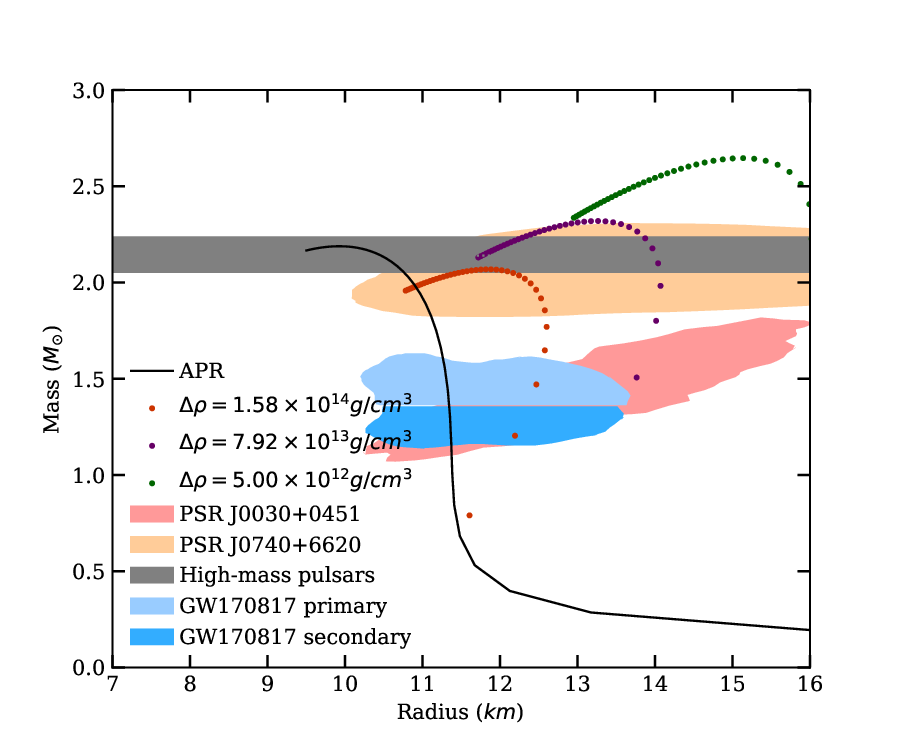}
\caption{\label{fig:MR_vary_drho} Similar to Fig.~\ref{fig:MR_vary_cs2} but varying $\Delta \rho$ with $(P_t, c_s^2) = (2\times10^{34}~\text{dyn cm}^{-2},0.48)$ (left) and $(P_t, c_s^2) = (2\times10^{33}~\text{dyn cm}^{-2},0.33)$ (right).
}
\end{figure}

Next, we focus on the effect of $\Delta \rho$ on the $M$--$R$ relations for HSs. Figure~\ref{fig:MR_vary_drho} presents such $M$-$R$ relations by increasing the density gap starting from the value used in Fig.~\ref{fig:MR_vary_cs2} up to $\sim2\times10^{14}~\text{g cm}^{-3}$, for the higher (lower) $P_t$ in the left (right) panel. This causes the $M$--$R$ curves to shift towards the lower left corner, resulting in a smaller maximum mass and radius. In other words, the density gap effectively softens the EOSs, indicated by the reduction in pressure at the same density (see Fig.~\ref{fig:EOS example}).

Combining the results in Figs.~\ref{fig:MR_vary_cs2} and \ref{fig:MR_vary_drho}, we conclude that the $M$-$R$ constraints favour HS EOSs with larger $c_s$ and smaller $\Delta \rho$ for high $P_t$. For the lower $P_t$ case, a larger $\Delta \rho$ with an intermediate $c_s$ is more likely to satisfy the constraints.

\subsection{Love--$C$ relations}

Let us next study the Love--$C$ relations for HSs, accounting for the elasticity in the QM core. 
For NSs, the relation can be fitted with a form \cite{Yagi_2017}
\begin{align}
    C = a_0 + a_1 \ln{\bar\lambda} + a_2 (\ln{\bar\lambda})^2, \label{eq:love_c_fit}
\end{align}
with $a_0 = 0.360$, $a_1 = -0.0355$, and $a_2 = 0.000705$  (see also \cite{Carson_2019} for a similar fit). 
The NS relation is found to be EOS-insensitive up to a 10\% variation~\cite{Yagi_2017}. In this section, we use APR for the NM EOS. In Appendix \ref{NL3}, we present the Love--$C$ relations for HSs with NL3 and MPa as the NM EOS and demonstrate that the results are qualitatively the same between the two EOSs.

\subsubsection{Dependence on the stiffness of quark matter}

We begin by studying how the Love--$C$ relations for HSs depend on $c_s$. 
For the tidal deformability calculation, we employ the shear modulus model described in Eq.~\eqref{eq:shear_modulus}, with the parameters $\mu_\mathrm{ref} = 3.13\times10^{34}~\text{dyn cm}^{-2} = 19.54~\text{MeV fm}^{-3}$, and $N = 0.5$. 

In Fig.~\ref{fig:LoveC_vary_cs2}, we show the Love--$C$ relations for elastic/fluid HSs with various $c_s^2$, together with the NS relation and its EOS variation that covers the EOSs studied in \cite{Postnikov_2001}. We observe that the deviation from the NS relation becomes larger as one decreases $c_s^2$. This is because, in such a case, HSs with a fixed mass generally have smaller radii (see Fig.~\ref{fig:MR_vary_cs2}), leading to smaller compactness and larger tidal deformability. In both high and low $P_t$ cases, the deviations from the NS case can be larger than the EOS variation. 

These deviations in the HS Love--$C$ relations from the NS case are compared with the measurement uncertainties from current multimessenger observations of NSs in Fig.~\ref{fig:LoveC_vary_cs2}. The light-blue rectangular region represents the estimate of the measurement error of $C$ and $\bar \lambda$. The error on $C$ for a 1.4~$M_\odot$ NS, $C = 0.159^{+0.025}_{-0.022}$, is obtained in\cite{Silva_2021} using pulsar observations with NICER, while that on the tidal deformability, $\bar\lambda = 190^{+390}_{-120}$, is obtained from GW170817 with LIGO/Virgo \cite{Abbott_2017a, Abbott_2018}. We observe that the deviations for the HS Love--$C$ relation from the NS case are much smaller than the current measurement uncertainties, making the two relations indistinguishable from current observations. Furthermore, elastic HSs giving larger deviations in the Love--$C$ relations tend to have a smaller maximum mass and are thus inconsistent with the presence of high-mass pulsars like J0740+6620. 

\begin{figure}[H]
\includegraphics[width=6.cm]{ 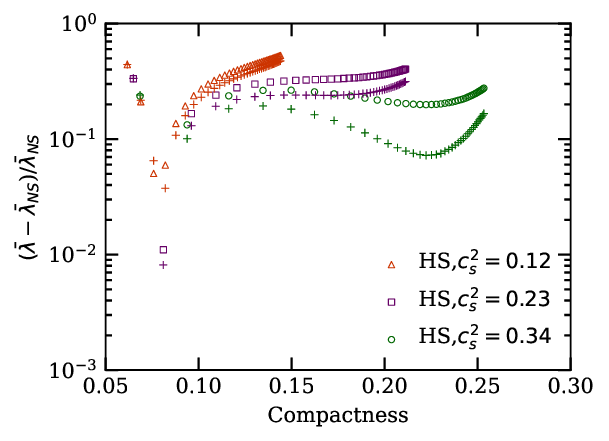}
\includegraphics[width=6.cm]{ 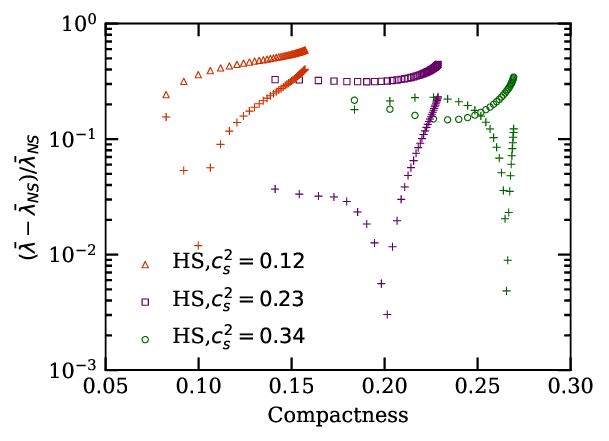}
\caption{\label{fig:Relative error} 
Fractional difference of the tidal deformability for HSs from the NS fit in~\cite{Yagi_2017} against compactness. Similar to Fig.~\ref{fig:LoveC_vary_cs2}, we vary $c_s^2$ and fix $P_t = 2\times10^{34}~\text{dyn cm}^{-2}$ (left) or $P_t = 2\times10^{33}~\text{dyn cm}^{-2}$ (right).
}
\end{figure}

To see these deviations more clearly, we show the fractional difference of the tidal deformability for HSs from the NS fit against compactness in Fig.~\ref{fig:Relative error}. In the low $P_t$ case (the right panel), the deviation in $\bar\lambda$ can be as large as 60\% (35\%) for elastic (fluid) HSs. For high $P_t$ (the left panel), they have deviations of similar size, around 50\% and 45\% for elastic and fluid HSs, respectively.

\subsubsection{Dependence on the energy density discontinuity}

As in Sec.~\ref{ssec:MR_constrants}, we next check how $\Delta \rho$ affects the Love--$C$ relations for the elastic HSs. 
Similar to the Earth's tides, a tidally deformed body in the fluid envelope can create a load that contributes to deforming the solid layer underneath. This fluid load acts against the tidal deformation of the solid core and reduces its overall tidal deformation. As a result, the presence of a dense fluid envelope is expected to effectively screen away the external tidal field acting on the core, reducing the effect of shear modulus on the overall tidal deformability \cite{Beuthe_2015_2, Lau_2019}. The effect of this screening depends on the thickness and the density of the envelope fluid layer and, therefore, is influenced by $\Delta \rho$.
In the following, we demonstrate that $\Delta \rho$ affects the Love--$C$ relations in a non-monotonic way.

\begin{figure}[H]
\includegraphics[width=6.cm]{ 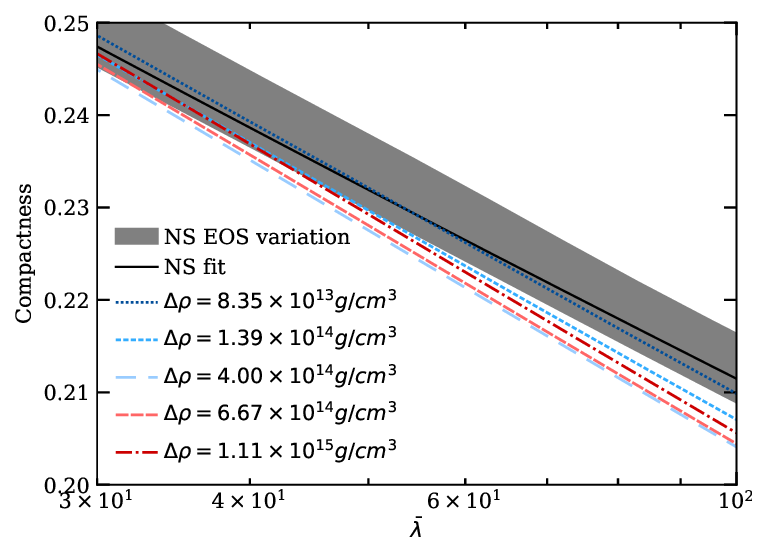}
\includegraphics[width=6.cm]{ 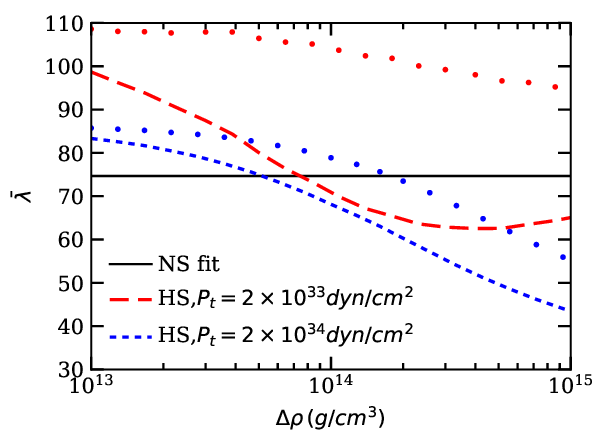}
\caption{\label{fig:Love_delta_rho} (Left panel) Love--$C$ relations of elastic HSs for various energy density gaps with fixed $(P_t,c_s^2) = (2\times10^{33}~\text{dyn cm}^{-2},0.48)$ and using APR for the NM EOS. The solid line and grey region are the same as Fig.~\ref{fig:LoveC_vary_cs2}. (Right panel) Relation between the dimensionless tidal deformability and the energy density gap for the high and low transition pressure with fixed $(C,c_s^2) = (0.22,0.48)$. We present the results for elastic (dashed) and fluid (dotted) HSs. The horizontal solid line represents the corresponding value of the dimensionless tidal deformability for the NS. In both panels, we use the shear modulus parameters as $(\mu_\mathrm{ref},N)=(3.13\times10^{34}~\text{dyn cm}^{-2},0.5)$.
}
\end{figure}

In the left panel of Fig.~\ref{fig:Love_delta_rho}, the Love--$C$ relations of five elastic HS models with various $\Delta \rho$ are compared with the fluid NS relations, showing certain deviations for large $\bar\lambda$. Notice that the HS model with the largest deviation has $\Delta \rho = 4\times10^{14}~\text{g cm}^{-3}$, which is the intermediate value among the five HS models. With such deviations, the Love--$C$ relations for elastic HSs can be distinguished from the NS relation whose EOS variation is shown by the grey band. As indicated in Fig.~\ref{fig:LoveC_vary_cs2}, however, this difference is much smaller than the current measurement uncertainties of $\bar\lambda$ and $C$ from multimessenger observations.

The right panel of Fig.~\ref{fig:Love_delta_rho} illustrates the $\bar\lambda$ dependence on $\Delta \rho$ for HS models with high and low $P_t$ values. $\bar\lambda$ of fluid HSs is also shown to study the effect of shear modulus.
For higher $P_t$, $\bar\lambda$ monotonically decreases as we increase $\Delta \rho$.
In the lower $P_t$ case, however, $\bar\lambda$ for elastic HSs decreases when $\Delta \rho$ increases from $10^{13}~\text{g cm}^{-3}$ to $10^{14}~\text{g cm}^{-3}$ but increases afterwards, denoting a non-montonic behavior. 
Comparing with the fluid HS models, we see that the effect of reduction in $\bar\lambda$ due to the solid QM core increases initially with $\Delta \rho$, and gradually saturates and starts decreasing at $\Delta \rho \sim 10^{14}~\text{g cm}^{-3}$.

Such a non-monotonic dependence on $\Delta \rho$ is not surprising and can also be seen in a simpler model within Newtonian gravity. Here, we consider a two-layer incompressible model with a solid core with uniform density $\rho_c$ and a fluid envelope of density $\rho_f = \rho_c-\Delta \rho$. The interface separating the two layers again has a transition pressure $P_t$. The solid core has a constant shear modulus $\mu_m$. The tidal deformability of this model has an analytical form \cite{Beuthe_2015_1, Beuthe_2015_2, Thesis} (see also Appendix~\ref{app:incompres}).

\begin{figure}[H]
\includegraphics[width=6.cm]{ 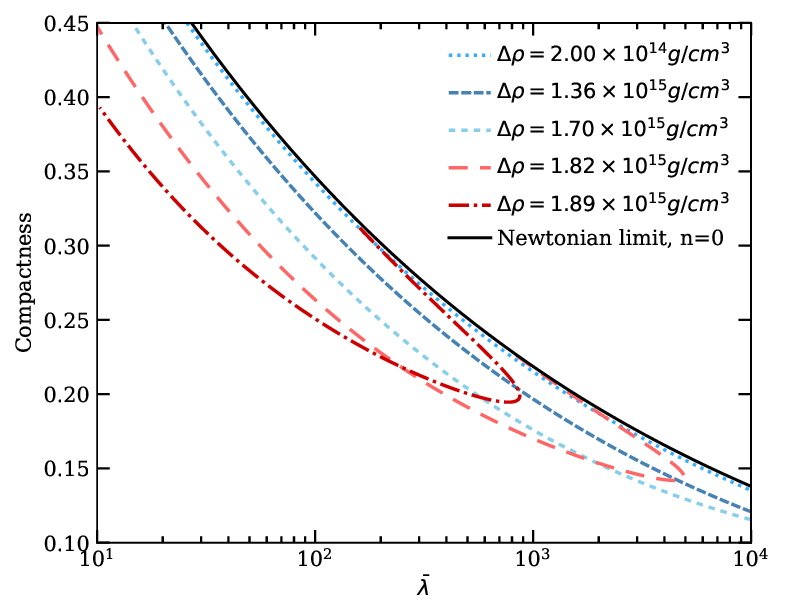}
\includegraphics[width=6.cm]{ 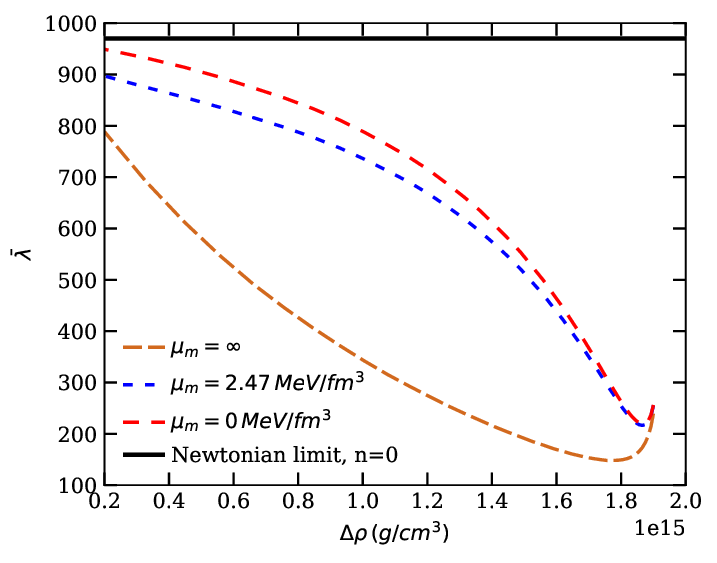}
\caption{\label{fig:LoveC_analytic} (Left panel) Newtonian Love--$C$ relation for the two-layer incompressible model. We choose the density of the core and the constant shear modulus as $\rho_c = 2\times10^{15}~\text{g cm}^{-3}$ and  $\mu_m = 2.47~\text{MeV fm}^{-3}$, respectively. The density of the envelope is adjusted by tuning the density gap $\Delta \rho$. The Newtonian Love--$C$ relation of a single-layer elastic incompressible star (where we set $y=1$ in Eq.~\eqref{eq:Love_analytic}), and fluid polytropic stars with polytropic index $n = 0$ are shown with a solid line.
(Right panel) Relation between $\bar\lambda$ and $\Delta \rho$ for the Newtonian two-layer incompressible star for various $\mu_m$ with fixed compactness of $C = 0.22$.
}
\end{figure}

The corresponding Love--$C$ relation for this two-layer incompressible model is illustrated in the left panel of Fig.~\ref{fig:LoveC_analytic}. As we increase $\Delta \rho$, the Love--$C$ relations progressively deviate towards smaller compactness and tidal deformability from the case for Newtonian incompressible stars with $ n=0 $,  $\bar\lambda = 1/(2 C^5)$ (see, e.g., \cite{Yagi_2013}). Eventually, a ``U-shaped'' Love--$C$ relation emerges, as represented by the last two color-coded models in this figure. This demonstrates that the dependence of the Love--$C$ relation on $\Delta \rho$ is not simply monotonic.

The non-monotonic dependence is further illustrated in the right panel of Fig.~\ref{fig:LoveC_analytic}, where we show $\bar\lambda$ versus $\Delta \rho$ for $\mu_m = 0$, $2.47~\text{MeV fm}^{-3}$, and $\infty$ at $C=0.22$ (we only show smaller $\bar \lambda$ if it takes two values at this compactness). As we increase $\Delta \rho$, $\bar\lambda$ decreases until it reaches a minimum at large $\Delta \rho$. When we further increase $\Delta \rho$, the ``U-shape'' behavior appears (as shown in the left panel of Fig.~\ref{fig:LoveC_analytic}). 

\begin{figure}[H]
\centering
\includegraphics[width=6cm]{ 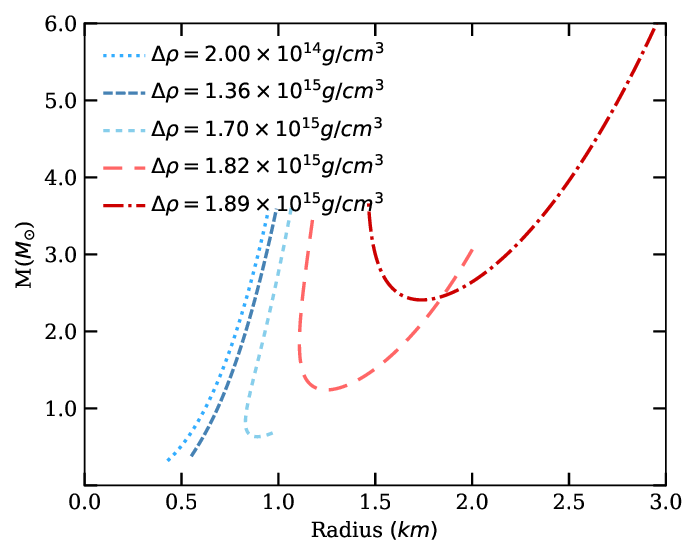}
\caption{\label{fig:MR_analytic} $M$--$R$ relation for the Newtonian two-layer incompressible models with various $\Delta \rho$.}
\end{figure}

Lastly, we show the $M$-$R$ relations of this analytical Newtonian model in Fig.~\ref{fig:MR_analytic}, which are quite different from the HS models. At smaller $\Delta \rho$, the relations are close to a simple power law with an index of $3$ (as expected from the case for single-layer incompressible stars, $M \propto R^3$), while the relations become non-monotonic for larger $\Delta \rho$. This also contributes to the ``U-shape'' behavior in Fig.~\ref{fig:LoveC_analytic}.

\subsubsection{Dependence on the shear modulus profile}

\begin{figure}[H]
\includegraphics[width=6.cm]{ 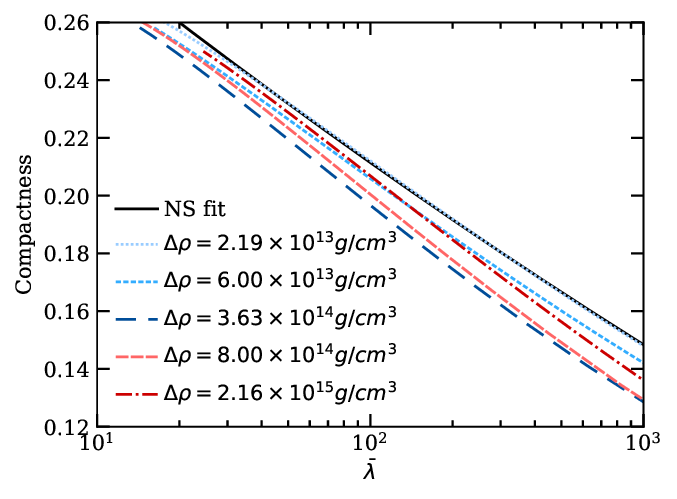}
\includegraphics[width=6.cm]{ 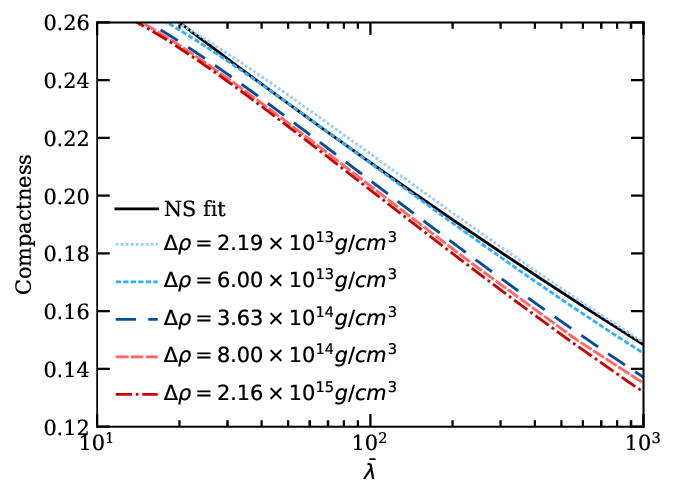}
\caption{\label{fig:love_C_vary_N}
Love--$C$ relations for elastic HSs with a shear modulus profile index of $N=0$ (left) and $N=1$ (right). We set other parameters as $(P_t,\Delta \rho, c_s^2) = (2\times10^{34}~\text{dyn cm}^{-2},3.13\times10^{34}~\text{dyn cm}^{-2}, 0.33)$ and use APR for the NM EOS. We also present the NS fit in Eq.~\eqref{eq:love_c_fit} for comparison.
}
\end{figure}

Lastly, we study the effect of the shear modulus profile index, $N$ (see Eq.~\eqref{eq:shear_modulus}). In the previous sections, we mainly adopted $N=0.5$, corresponding to the high-density limit of the shear modulus of the CCS phase. Here, we also consider the cases $N=0$ (constant shear modulus) and $N=1$ (linear shear modulus). Fig.~\ref{fig:love_C_vary_N} presents the Love--$C$ relations for elastic HSs for these choices of $N$.

For $N=0$, the deviation from the NS case once again has a non-monotonic behavior. The deviation at first increases as one increases $\Delta \rho$, but it starts to decrease after $\Delta \rho \sim 3 \times 10^{14} ~\text{g cm}^{-3}$. On the other hand, for $N=1$, both compactness and tidal deformability decrease monotonically as one increases $\Delta \rho$. This demonstrates a different dependence of the Love--$C$ relation on $\Delta\rho$ for different shear modulus profiles. Nevertheless, the maximum deviation of the two cases is of similar size.

\section{Conclusions}
\label{sec:conclusion}

In this paper, we studied the effect of a solid QM core on HS observables via EM and GW measurements. We constructed the QM part of the HSs using the CSS EOS and the NM part with realistic nuclear EOSs. These two phases were separated by a sharp phase transition with a density discontinuity. The shear modulus of the QM, which follows a power-law relation with density, was assumed to have no effect on the static spherically symmetric configuration, i.e., only affecting the perturbed quantities such as the tidal deformability.

We first compared the $M$-$R$ relations of the HS models with the observational constraints from EM and GW measurements. 
We found the maximum mass and the corresponding radius of the HS models to increase with the stiffness of the QM core, parametrized by the speed of sound ($c_s$). Meanwhile, a higher density discontinuity ($ \Delta \rho $) effectively softened the EOS.

We then considered the Love--$C$ relations for elastic HSs. 
We numerically solved for $\bar\lambda$ of elastic HSs using the perturbation equations in \cite{Lau_2019, Gittins_2020} to obtain the Love--$C$ relations. 
The presence of a solid core generally caused $\bar\lambda$ to be lower than fluid HSs. 
We found the following regarding the dependence of the Love--$C$ relations on the QM EOS parameters:
\begin{enumerate}
    \item dependence on the speed of sound $c_s$: The Love--$C$ relations of elastic HSs showed significant deviations from the NS relation in contrast to fluid HSs. The deviations were larger for models with low $P_t$.
In particular, some models showed deviations up to 60\%, while those with a fluid core only had deviations by less than 35\%.
\item dependence on the energy density gap $\Delta \rho$: Deviations in the Love--$C$ relations for elastic HSs from the NS case showed a non-monotonic behavior in $\Delta \rho$.  To reinforce our results, we demonstrated a similar non-monotonic behavior of the Love--$C$ relations in Newtonian two-layer incompressible models with a solid core.
\item dependence on the index $N$ for the shear modulus profile: We found the maximum deviation is of similar size for $N = 0, 0.5, 1$, but the dependence of the Love--$C$ relations on $\Delta \rho$ is affected by the value of $N$.
\end{enumerate}

Our results showed substantial deviations in the Love--$C$ relations for elastic HSs from the fluid NS case when the shear modulus is large, the transition pressure is low, and the density gap is large. However, the deviations are unlikely to be detectable from existing NS observations. Moreover, the deviations in the Love--$C$ relations are suppressed if we restrict the QM EOS parameter space to satisfy the constraints from the mass-radius measurements. Thus, it may be challenging to probe elastic HSs with the Love--$C$ relations alone, even with future NS observations. One can still expect, however, to distinguish HSs and NSs with future GW observations from the improved measurement of the tidal deformability. For example, the recent work by \cite{Mondal_2023} has shown the possibility of detecting the presence of a strong first-order phase transition with future GW detectors. This can be further extended to investigate the measurability of the elasticity in HSs with solid QM cores.

Note again that the above conclusion on the Love-$C$ relations is based on the ``slow conversion condition" at the QM-NM interface under perturbations. We provide a brief discussion on the other extreme where the QM-NM conversion happens immediately under perturbations in Appendix~\ref{app:rapid_bc}. The resulting deviations of the Love-$C$ relations of elastic HSs from that of the fluid NSs are significantly smaller in this case.

\begin{acknowledgements}
K.Y. acknowledges support from NSF Grant PHYS-2339969 and the Owens Family Foundation.
\end{acknowledgements}

\appendix

\section{Elastic HSs with NL3 or MPa for NM EOS}
\label{NL3}
To see how our main results change if we use a different NM EOS, in this appendix, we consider NL3 and MPa as the NM EOS, which exhibit different characteristics in stiffness compared to APR. As we will see below, these EOSs do not satisfy the bounds from GW170817. Therefore, the results presented in this appendix should only be considered as a reference and not be taken seriously from the observation viewpoint.

\begin{figure}[H]
    \centering
    \subfigure[]{
        \includegraphics[width=6.cm]{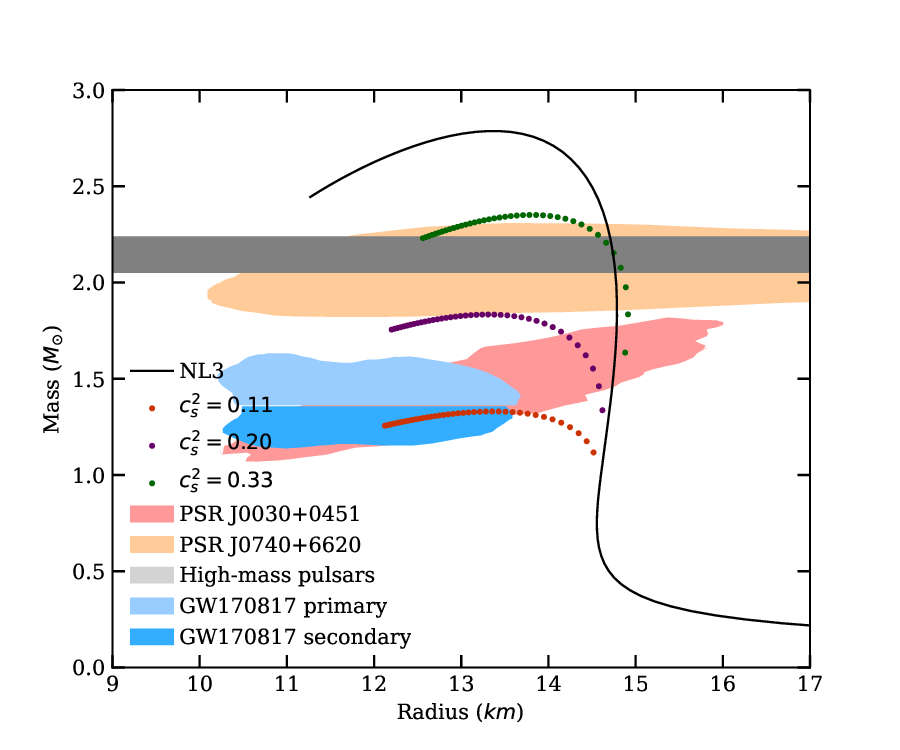}
        \includegraphics[width=6.cm]{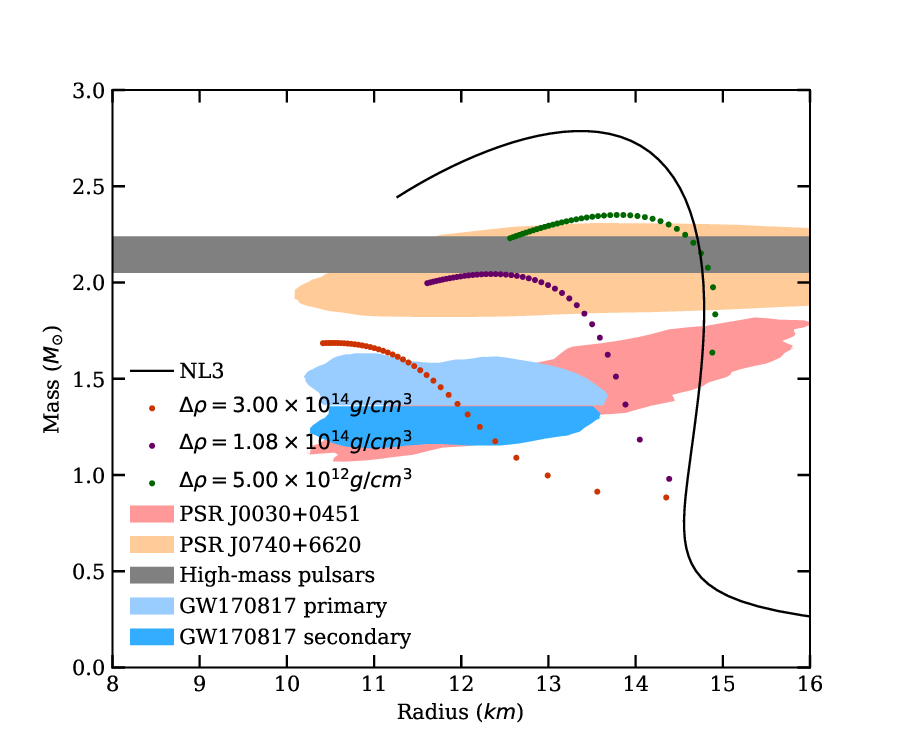}
    }\\[-1.9ex]
    \subfigure[]{
        \includegraphics[width=6.cm]{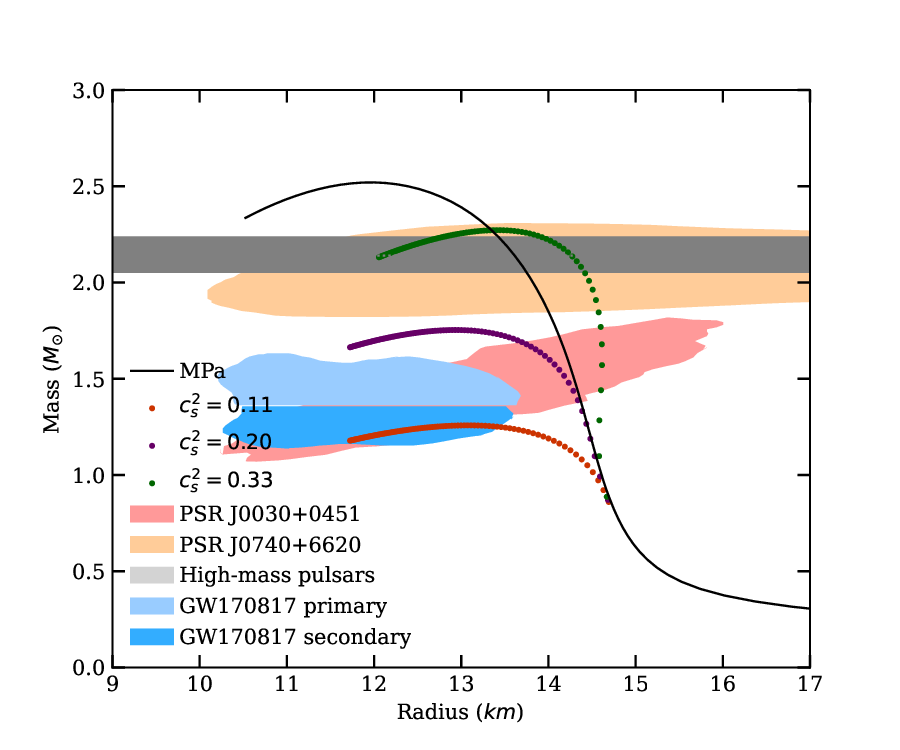}
        \includegraphics[width=6.cm]{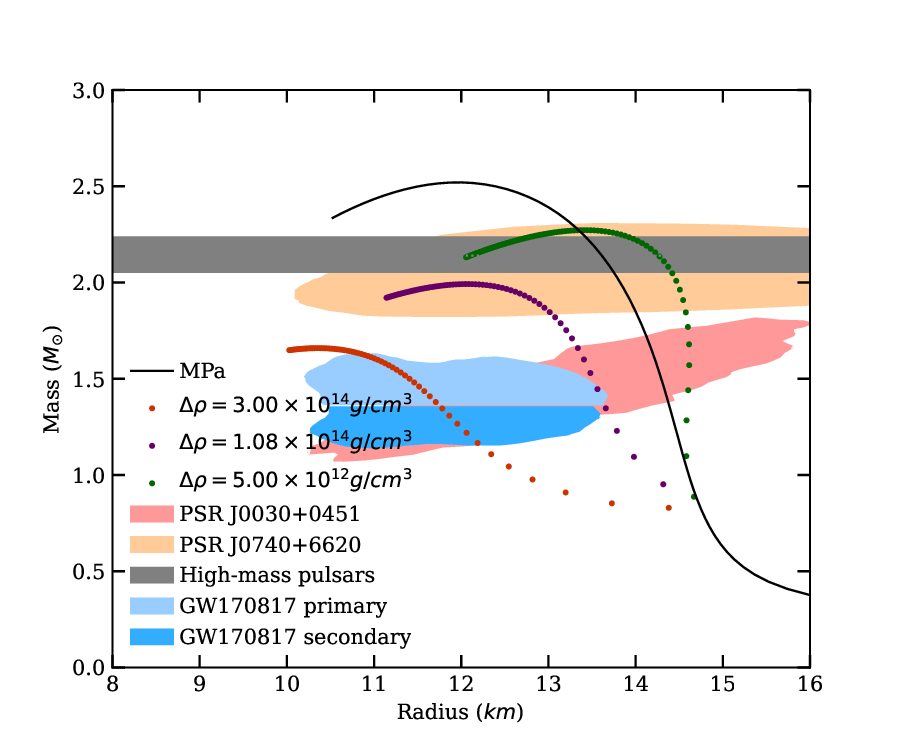}
    }
    \caption{\label{fig:NL3&MPa_MR} Similar to Figs.~\ref{fig:MR_vary_cs2} and \ref{fig:MR_vary_drho} but with NL3 (a) and MPa (b) for the NM EOS. We vary $c_s^2$ with $\Delta \rho = 5\times10^{12}~\text{g cm}^{-3}$ (left) and vary $\Delta \rho$ with $c_s^2 = 0.33$ (right). In both panels, we fix $P_t = 2\times10^{34}~\text{dyn cm}^{-2}$.}

\end{figure}

Figure~\ref{fig:NL3&MPa_MR} presents the $M$--$R$ relations for HSs with NL3 (a) and MPa (b). Similar to those in Sec. \ref{ssec:MR_constrants} with APR, the stellar mass increases as one increases $c_s^2$ and decreases $\Delta \rho$. Both NL3 and MPa are stiffer EOSs than APR, leading to a higher maximum mass and larger radius for HSs. MPa is slightly softer than NL3 due to the effects of hyperons at high densities. Notice that most astrophysical bounds can be satisfied for HSs with these stiff NM EOSs if $c_s^2$ is larger or $\Delta \rho$ is small, except for the bounds from GW170817, which prefer softer EOSs.

\begin{figure}[H]
\includegraphics[width=6.cm]{ 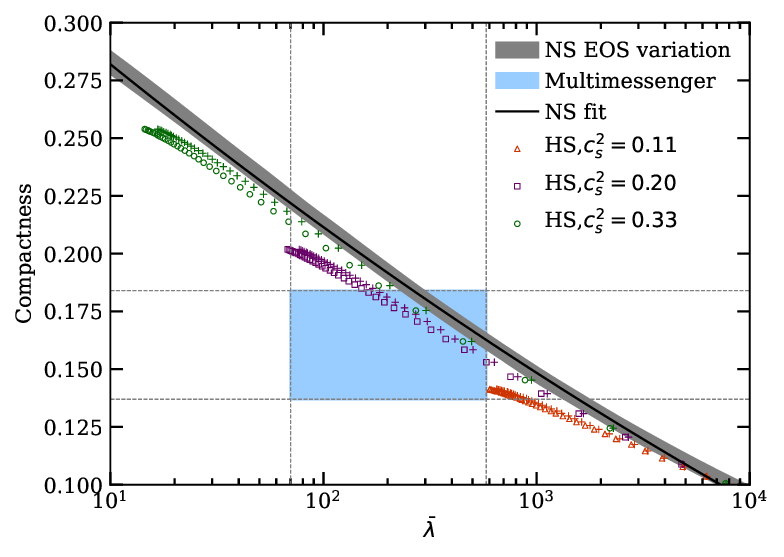}
\includegraphics[width=6.cm]{ 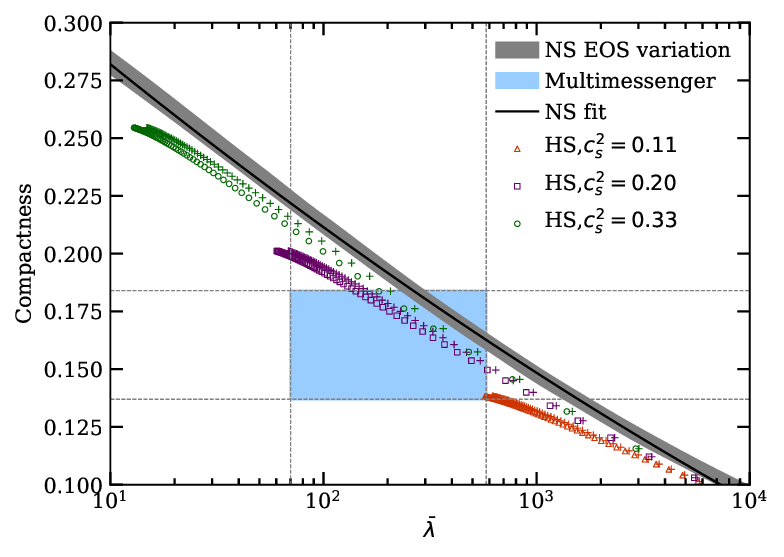}
\caption{\label{fig:NL3&MPa_LC} Similar to  Fig.~\ref{fig:LoveC_vary_cs2} but with NL3 (left) and MPa (right) as the NM EOS. We vary $c_s^2$ while fix $(P_t,\Delta \rho) = (2\times10^{34}~\text{dyn cm}^{-2},1.08\times10^{14}~\text{g cm}^{-3})$. We use the shear modulus parameters as $(\mu_\mathrm{ref},N)=(3.13\times10^{34}~\text{dyn cm}^{-2},0.5)$.}
\end{figure}

Figure~\ref{fig:NL3&MPa_LC} shows the Love--$C$ relations for elastic/fluid HSs with NL3 (left) and MPa (right). Despite the difference in the $M$--$R$ relations, the Love--$C$ relations are very similar to those in the left panel of Fig.~\ref{fig:LoveC_vary_cs2} for HSs with APR for similar choices of EOS parameters. These results demonstrate that our main results with APR are qualitatively valid even for other NM EOSs.

\section{Tidal Perturbation Formalism}
\label{sec_formalism}

This appendix summarizes the formalism in \cite{Lau_2019} for the time-independent perturbation problem of the solid core of an HS. This formalism is rewritten from the one in \cite{Penner_2011} after correcting some typos. We also note that \cite{Pereira_2020, Gittins_2020} also provided the amended formalism with the same choice of dependent variables as \cite{Penner_2011}, which we have verified to be consistent with our equations.

Let us define some new perturbation functions.
Following \cite{Finn_1990, Kruger_2015}, the strain tensor in Eq.~\eqref{eq:Hookes} contains two independent components, which can be expressed in terms of the perturbed metric and Lagrangian displacement through Eq.~\eqref{eq:strain_tensor}:
\begin{align}
    S_r(r) Y_{lm} &\equiv \delta \Sigma_r^r = \frac{1}{3} \left[-K(r)+H_2(r) + \frac{l(l+1)}{r^2}V(r) \right. \nonumber \\
    &\qquad\qquad\qquad\qquad\quad \left. + \frac{2}{r} \frac{dW(r)}{dr} - \left(\frac{4}{r^2} - \frac{u^\prime}{r}\right)W(r)\right] Y_{lm}, \\
    S_\perp(r) \partial_A Y_{lm} &\equiv \delta \Sigma_A^r = \frac{e^{-u}}{2 r} \left[\frac{dV(r)}{dr} - \frac{2V(r)}{r} + e^{u}\frac{W(r)}{r}\right]\partial_A Y_{lm},
\end{align}
where the index $A$ runs between $\theta$ and $\phi$.

In our formalism, we introduce the stress variables $Z_r$, and $Z_\perp$, denoting the Lagrangian perturbation of the stress in the radial and tangential directions, respectively:
\begin{align}
Z_r(r) \equiv& \Delta p(r) - 2 \mu S_r(r), \label{Z_r}\\
Z_\perp(r) \equiv& -2 \mu S_\perp(r),\label{Z_t}
\end{align}
where $\Delta p(r)$ is the radial component of the expansion of the Lagrangian pressure perturbation in terms of spherical harmonics.
We also define the variable $J$ as
\begin{equation}
J(r) \equiv H_0^\prime(r) - 8\pi e^u (\rho + p) \frac{W(r)}{r} + 16 \pi \nu^\prime \mu V(r), \label{eq_J_def}
\end{equation}
which replaces $H_0^\prime$ as one of the dependent variables in our formalism. This quantity is continuous across the density discontinuity at the QM-NM transition in the HS, unlike $H_0^\prime$ or $V$, as long as $W$ is continuous.

The complete set of equations governing the perturbations is given by \cite{Lau_2019}
\begin{align}
%equation 1================================================
\frac{d W}{dr} =& \Big(1 - \frac{2 \alpha_2}{\alpha_3} - \frac{r u^\prime}{2}\Big)\frac{W}{r} - \frac{r}{\alpha_3} Z_r + \frac{\alpha_2}{\alpha_3}l(l+1) \frac{V}{r} - \frac{\alpha_2}{\alpha_3} r K - \frac{1}{2} r H_2, \label{fn_eqt1} \\
%equation 2================================================
\frac{d Z_r}{dr} =& \Bigg[ p^\prime \bigg( \frac{r \nu^{\prime \prime}}{\nu^\prime} - \frac{r u^\prime}{2} - 2 \bigg) - \frac{4}{r} \frac{\alpha_1 }{\alpha_3} \big(\alpha_3 + 2 \alpha_2 \big)  \Bigg] \frac{W}{r^2} \nonumber \\
& - \bigg( \frac{r \nu^\prime}{2} + \frac{4 \alpha_1}{\alpha_3}\bigg) \frac{Z_r}{r} +\bigg[p^\prime + \frac{2}{r}\frac{\alpha_1}{\alpha_3}\Big(\alpha_3 + 2 \alpha_2\Big) \bigg]\frac{l(l+1) V}{r^2} + \frac{l(l+1)}{r} e^u Z_\perp \nonumber\\
& + \frac{1}{2}(\rho + p) H_0^\prime -  \Big[p^\prime + \frac{2}{r} \frac{\alpha_1 }{\alpha_3} \big(\alpha_3 + 2 \alpha_2 \big) \Big]K - \frac{p^\prime}{2}H_2, \label{fn_eqt2}\\
%equation 3================================================
\frac{d V}{dr} =& -e^u \frac{W}{r} + \frac{2 V}{r} -\frac{r e^u}{\alpha_1} Z_\perp, \label{fn_eqt3}\\
%equation 4================================================
\frac{d Z_\perp}{dr} =& \bigg[p^\prime + \frac{2}{r}\frac{\alpha_1}{\alpha_3}\big(\alpha_3 + 2 \alpha_2\big)\bigg]\frac{W}{r^2} - \frac{\alpha_2}{\alpha_3} \frac{Z_r}{r} \nonumber\\
& + \bigg[\frac{2 \alpha_1}{r} - 2 \alpha_1 \Big(1 + \frac{\alpha_2}{\alpha_3}\Big)\frac{l(l+1)}{r}\bigg] \frac{V}{r^2} - \bigg[\frac{r u^\prime}{2} + \frac{r \nu^\prime}{2} + 3\bigg] \frac{Z_\perp}{r} \nonumber\\
& +\frac{1}{2}(\rho + p) \frac{H_0}{r} + \frac{\alpha_1}{\alpha_3} \Big(\alpha_3 + 2 \alpha_2 \Big) \frac{K}{r}, \label{fn_eqt4}\\
%equation 5================================================
\frac{d H_0}{d r} =& J + \frac{1}{r^2}(\nu^\prime + u^\prime) W - 16 \pi \alpha_1 \nu^\prime V, \label{fn_eqt5}\\
%equation 6================================================
\frac{d J}{d r} =& \bigg[\frac{32 \pi e^u}{r^2} \frac{ \alpha_1}{\alpha_3}\big(\alpha_3 + 2 \alpha_2\big) -\frac{3}{2 r^2}\nu^\prime\big(u^\prime + \nu^\prime\big)\bigg] W - 8 \pi e^u \frac{1}{\alpha_3}\Big( \alpha_3 + 2 \alpha_2\Big)Z_r \nonumber\\
& -\frac{8 \pi}{r^2}\Big[\big(\rho + p \big)e^u l(l+1) + \frac{2 \alpha_1}{\alpha_3}\big(\alpha_3 + 2 \alpha_2\big) e^u l(l+1) + 4 \alpha_1 \big(1 - e^u\big) - 2\alpha_1 \big(r \nu^\prime\big)^2\Big] V \nonumber\\
& - 16 \pi e^u \big(r \nu^\prime\big)Z_\perp + \Big[l(l+1) e^u + 2\big(e^u - 1\big) - r\big(\frac{u^\prime}{2} + \frac{5 \nu^\prime}{2}\big) + \big(r \nu^\prime\big)^2\Big]\frac{H_0}{r^2} \nonumber\\
& + \Big[\frac{r}{2}\big(u^\prime - \nu^\prime \big) - 2\Big] \frac{J}{r} +  \Big[\frac{1}{r}\big(u^\prime + \nu^\prime\big) + 16 \pi e^u \frac{\alpha_1}{\alpha_3}\big(\alpha_3 + 2\alpha_2\big)\Big]K. \label{fn_eqt6}
\end{align}

The above equations are formulated in a way to be compared with the Newtonian linear perturbation problems (e.g., \cite{Saito_1974}). The quantities $\alpha_1$, $\alpha_2$, $\alpha_3$ are defined to represent different elastic moduli in an isotropic solid:

\begin{equation}
\alpha_1 = \mu, \qquad \alpha_2 = c_s^2 (\rho + p) - \frac{2}{3}\mu, \qquad \alpha_3 = c_s^2 (\rho + p) + \frac{4}{3}\mu,
\end{equation}
where $\alpha_2$ and $\alpha_3$ represent the relativistic generalization of the Lam\'e coefficient and the P-wave modulus defined in linear elastic theory, respectively (see e.g., \cite{Landau_1984, Mavko_2003}). The (equilibrium) speed of sound $c_s$ is defined in Eq.~\eqref{eq:cs}.

The variable $K$ is expressed in terms of the new set of dependent variables by combining the ($r$, $r$) \& ($r$, $\theta$) components of the Einstein equations:

\begin{align} \label{fn_alg1}
(l+2)(l-1) \; e^u K =& - 16\pi e^u r^2 Z_r - 16 \pi e^u \Big(2 + r \nu^\prime \Big) r^2 Z_\perp \nonumber\\
&+ \left[l(l+1) e^u - 2 + \big(r \nu^\prime \big)^2 \right] H_0 + r^2 \nu^\prime J.
\end{align}

Equations~\eqref{fn_eqt1}--\eqref{fn_eqt4} can be derived solely from the energy-momentum conservation, while the remaining two are obtained from the Einstein equations.

Equations~\eqref{fn_eqt1}--\eqref{fn_eqt6}, together with the algebraic relation for $K$ (Eq.~(\ref{fn_alg1})) form a set of six coupled ODEs with six dependent variables $\Big(W$, $Z_r$, $V$, $Z_\perp$, $H_0$, $J\Big)$. Note that $H_2$ is algebraically related to other variables through
\begin{align}
    H_2 = H_0 + 32\pi \mu V.
\end{align}

The perturbations of the fluid envelope are described by a second-order ODE in $H_0$ \cite{Hinderer_2008, Hinderer_2009, Damour_2009}, which correspond to the $\mu\rightarrow 0$ limit of Eqs.~\eqref{fn_eqt1}-\eqref{fn_eqt6}:
\begin{align}
    H_0^{\prime\prime} &+ \left\{\frac{2}{r} + e^u\left[\frac{2m}{r} + 4\pi r(p-\rho)\right]\right\} H_0^{\prime} \nonumber \\
    &+ \left[-\frac{l(l+1)}{r^2} + 4\pi e^u\left( 5\rho + 9 p + \frac{\rho+p}{c_s^2}\right) - {\nu^\prime}^2\right] H_0 = 0. \label{eq:fluid_H0}
\end{align}

\section{Computation method}\label{Computation method}

Let us next describe a procedure to solve Eqs.~\eqref{fn_eqt1}--\eqref{fn_eqt6} numerically.
We start integrating these equations from the stellar center towards the solid-fluid interface. By expanding the variables to the second order in $r$ about $r=0$, the dependent variables are written in the form \cite{Finn_1990}
\begin{align}
    Y(r) = Y^{(0)} r^n + Y^{(2)} r^{n+2} +\mathcal{O}(r^{n+4}),
\end{align}
where $n = l$ for ($W$, $V$, $H_0$), $l-2$ for ($Z_r$, $Z_\perp$) and $(l-1)$ for ($J$). Here we consider the quadrupolar tidal perturbation with $l=2$. Inserting this ansatz into Eqs.~\eqref{fn_eqt1}--\eqref{fn_eqt6}, we obtain nine independent constraints whose explicit forms are given by Eqs.~(A4)--(A12) in \cite{Lau_2019}. This leaves us with three independent regular solutions.

The Israel junction conditions require the continuities of  $\left(W, Z_r, Z_\perp, H_0, J\right)$ across the solid-fluid interface \cite{Israel_1966, Poisson_2009, Penner_2011, Lau_2019}. 
With three independent solutions within the solid, we need to impose two constraints from interface boundary conditions such that there remains one independent solution in the fluid layer.
Since $Z_\perp = 0$ in the fluid layer, the continuity condition requires $Z_\perp$ to vanish at the solid side of the interface, providing one constraint. The other constraint can be derived from the continuity of $W$, $Z_r$, and $H_0$. Explicitly, we write the constraints as
\begin{align}
    Z_r^{(\rm s)} &= Z_r^{(\rm f)} = \frac{\rho^{(\rm f)}+p}{2}\left(H_0 - \nu^\prime \frac{W}{R_c}\right), \label{eq:bdy_Zr}\\
    Z_\perp^{(\rm s)} &= 0,
\end{align}
where the quantities are evaluated at $r=R_c$, with $R_c$ being the radius of the solid core. We use superscripts (s) and (f) to indicate the solid or fluid side value of the quantity if it is discontinuous.

The dimensionless tidal deformability, $\bar\lambda$, can then be computed as follows.
The above two constraints allow us to determine the solution within the solid core up to an arbitrary constant factor, which comes from the fact that the amplitude of the perturbations has not been specified. We then use the continuity of $\left(W, H_0, J\right)$ to determine the values of $H_0$ and $H_0^\prime$ on the fluid side of the interface. In particular, using Eq.~\eqref{eq_J_def},
\begin{align}
{H_0^\prime}^{(\rm f)} = J + 8\pi e^{u}(\rho^{(\rm f)} + p) \frac{W}{R_c}. \label{eq:bdy_J}
\end{align}

In the fluid envelope, we integrate Eq.~\eqref{eq:fluid_H0} towards the stellar surface. The metric perturbations are matched with the exterior solution. The dimensionless tidal deformability of $l=2$, $\bar\lambda$, is then given by 
\begin{align}
    \bar\lambda =& \frac{16}{15}(1-2C)^2 \left[2 + 2C(y - 1) - y\right] \Bigg\{2C\bigg[4(1+y)C^4  \nonumber\\ 
&+ (6y-4)C^3+ (26-22y)C^2+ 3C(5y-8)	-3y+6 \bigg] \nonumber\\
&+ 3(1-2C)^2\big[2 - y + 2C (y - 1)\big] \text{log}(1 - 2C)\Bigg\}^{-1}, \label{eq:lambda_bar}
\end{align}
where $y = R H_0^\prime(R) / H_0(R) $.

When solving for the tidal deformability of a fluid HS, we integrate Eq.~\eqref{eq:fluid_H0} near $r=0$ with the regularity condition
\begin{align}
    H_0 = a r^l + \mathcal{O}(r^{l+2}),
\end{align}
for some arbitrary constant $a$. Across the interface, we have the junction condition
\begin{align}
    H_0^{(+)} =& H_0^{(-)}, \\
    H_0^{\prime(+)} =& H_0^{\prime(-)} + \frac{4\pi R_c^2}{m + 4\pi R_c^3 P_t}\left(\rho^{(+)}-\rho^{(-)}\right) H_0,
\end{align}
where the quantities with superscripts $(+)$ and $(-)$ are evaluated at $R_c+\delta$ and $R_c-\delta$ respectively for $0<\delta\ll R_c$. The quantities without the subscripts are continuous across the interface. $\bar \lambda$ can then be determined from Eq.~\eqref{eq:lambda_bar}.

\section{\label{app:rapid_bc} Rapid conversion in phase transition}
In this paper, we focus on the scenario in which the Lagrangian displacement is always continuous in the radial direction in the perturbed configuration, i.e., $W$ is continuous across the interface. An underlying assumption for this condition is that the phase conversion timescale is much longer than that of the perturbation process, e.g., the tidal forcing during a binary inspiral. This condition has been assumed in some previous studies \cite{Finn_1990, Penner_2011}.
Another possibility is that the phase conversion happens fast enough such that part of the perturbed matter near the interface undergoes a phase transition, resulting in a different junction condition \cite{Karlovini_2004, Pereira_2018, Tonetto_2020}:
\begin{align}
    W^{(\rm s)} + \frac{R_c Z_r^{(\rm s)}}{{p^\prime}^{(\rm s)}} = W^{(\rm f)} + \frac{R_c 
 \Delta p^{(\rm f)}}{{p^\prime}^{(\rm f)}}. \label{eq:rapid_bc}
\end{align}
Equations~\eqref{eq:bdy_Zr} and \eqref{eq:bdy_J} are then replaced by
\begin{align}
    Z_r^{(\rm s)} =& \frac{\rho^{(\rm s)}+p}{2}\left(H_0 - \nu^\prime \frac{W^{(\rm s)}}{R_c}\right), \\
{H_0^\prime}^{(\rm f)} =& J + 8\pi e^{u}(\rho^{(\rm f)} + p) \frac{W^{(\rm f)}}{R_c} \nonumber\\
=& J + 8\pi e^{u}(\rho^{(\rm f)} + p) \frac{W^{(\rm s)}}{R_c} - \frac{16\pi e^u}{\nu^\prime}\left(1-\frac{\rho^{(\rm f)}+p}{\rho^{(\rm s)}+p}\right)Z_r\nonumber\\
=& J -\frac{16\pi e^u}{\nu^\prime}Z_r^{(\rm s)} + \frac{8\pi e^u}{\nu^\prime} \left(\rho^{(\rm f)}+p\right)H_0.
\end{align}
Note that the junction conditions at the interface for $H_0$ and $H_0^\prime$ are the same for fluid HSs, and therefore the tidal deformability is the same for the slow conversion and rapid conversion scenarios for such HSs.

\begin{figure}[H]
\includegraphics[width=6.cm]{ 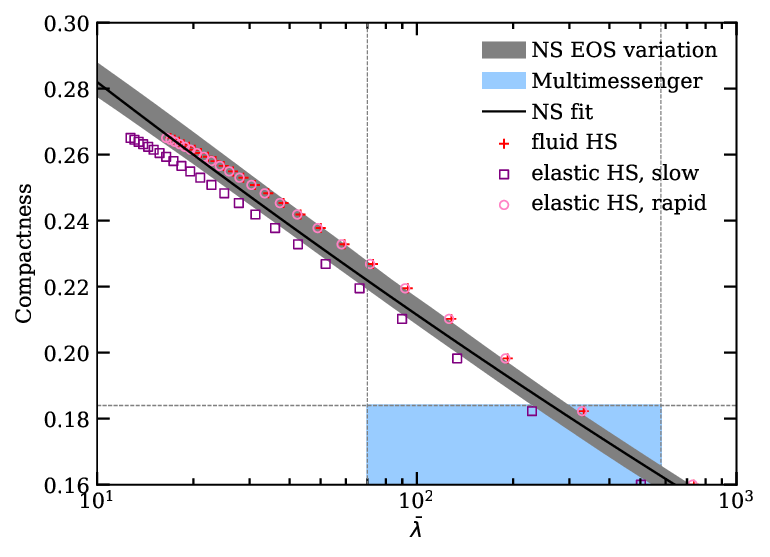}
\includegraphics[width=6.cm]{ 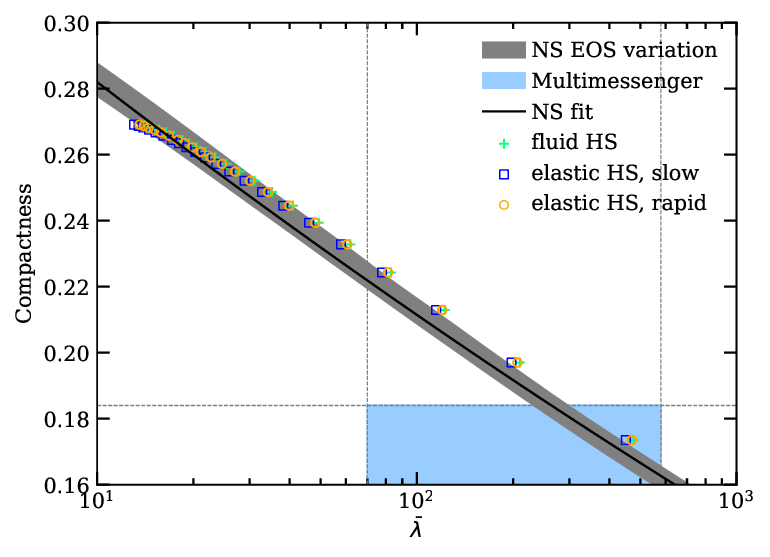}
\caption{\label{fig:BC} 
Love--$C$ relations for HSs with different boundary conditions. 
We use the same EOS parameters as the right panel of Fig.~\ref{fig:MR_vary_drho}, i.e., $(P_t, c_s^2) = (2\times10^{33}~\text{dyn cm}^{-2},0.33)$ and APR EOS.
We also choose $\Delta \rho = 7.92\times10^{13}~\text{g cm}^{-3}$ (left) and $\Delta \rho = 5\times10^{12}~\text{g cm}^{-3}$ (right). The black solid lines represent the NS fit, and the grey-shaded regions indicate the NS EOS variation. The blue-shaded regions correspond to the multimessenger constraints. The data points represent different models: fluid HS (crosses), elastic HS in slow conversion (squares), and elastic HS in rapid conversion (open circles).
}
\end{figure} 
In Fig.~\ref{fig:BC}, we compare the Love--$C$ relation of the original (slow) junction condition to the rapid condition for two different elastic HS models. The deviations from the fluid HSs are significantly smaller for the rapid conversion condition. This suggests the overall shear strain of the solid core is reduced under this condition, and hence the effect of elasticity is smaller. 
Also, as seen from Eq.~\eqref{eq:rapid_bc}, the discontinuity in $W$ is proportional to the difference of the inverse of the density at the two sides of the interface. Hence, the deviation in Love--$C$ relations between the slow and rapid conditions is much smaller in the right panel with a smaller $\Delta \rho$.
A thorough study of the condition at the phase transition is required to reveal the effect of the reaction rate on the tidal deformability, which we leave for future work.

\section{Newtonian Two-layer incompressible model} \label{app:incompres}

To understand the non-trivial dependence of the Love--$C$ relation on the density gap, we consider a simpler but related model consisting of a solid core and a fluid envelope, both of constant densities, in the Newtonian limit. Analytic expression exists for both $\bar\lambda$ and $C$.

For a two-layer incompressible model with a core density $\rho_c$, a core shear modulus $\mu_m$, a central pressure $p_c$, a transition pressure $P_t$, and an envelope composed of fluid with density $\rho_f = \rho_c - \Delta \rho$, the tidal deformability is given by \cite{Beuthe_2015_1, Beuthe_2015_2, Thesis}
\begin{align}\label{eq:Love_analytic}
\bar\lambda = \frac{1}{2 C^5} \Bigg\{1 + \cfrac{\quad (1- y) \; \Big[n_0(y) + n_1(y) \hat{\rho} + n_2(y) \hat{\rho}^2\Big] + \dfrac{19}{2} y^4 \hat{\mu}_c \qquad}{\qquad y^5 + \dfrac{2}{5} (1- y)\Big[ n_0(y) \hat{\rho} -  n_2(y) \hat{\rho}^2\Big] + \dfrac{19}{5} \hat{\gamma} y^4 \hat{\mu}_c \qquad}\Bigg\}^{-1},
\end{align}
where 
\begin{align}
y =& \frac{R_c}{R},\\
\hat\rho =& \frac{R^3 \rho_f }{M},\\
\hat{\gamma} =& \frac{\hat{\rho}}{1 - \hat{\rho}},\\
n_0(y) =& 1+y+y^2+y^3+y^4,\\
n_1(y) =& -\frac{1}{2} \; \Big(4+4 y+4 y^2-y^3-y^4\Big),\\
n_2(y) =& \frac{1}{2} \; \Big(2+2 y+2 y^2-3 y^3-3 y^4\Big).
\end{align}
Here, $R_c$ is the radius of the core, which corresponds to $p(R_c) = P_t$.

\bibliographystyle{spphys} 
\bibliography{master}

\end{document}